\def\HAL{1}

\if\HAL1

\documentclass[a4paper, 11pt]{article}
\usepackage{authblk}

\else

\documentclass[AMA,LATO1COL]{WileyNJD-v2}
\newcommand\BibTeX{{\rmfamily B\kern-.05em \textsc{i\kern-.025em b}\kern-.08em
T\kern-.1667em\lower.7ex\hbox{E}\kern-.125emX}}

\articletype{Article Type}%

\received{<day> <Month>, <year>}
\revised{<day> <Month>, <year>}
\accepted{<day> <Month>, <year>}

\fi

\usepackage{moreverb}

\usepackage{amsfonts}
\usepackage{textcomp}
\usepackage{xspace}
\usepackage{multirow}

\if\HAL1
\usepackage{algorithm, algorithmicx, algpseudocode}
\usepackage{listings}
\fi
\usepackage{graphicx}
\graphicspath{{pdf/}{jpeg/}{png/}{Figures/}}
\DeclareGraphicsExtensions{.pdf,.jpeg,.png}
\usepackage[tight,footnotesize]{subfigure}
\usepackage{url}
\usepackage{mathtools}

\usepackage{tikz}
\usetikzlibrary{calc}
\newcommand{\tikzmarkin}[1]{\tikz[overlay,remember picture] \node (#1) {};}
\newcommand{\tikzmarkout}[1]{\tikz[overlay,remember picture] \node (#1) {};}

\if\HAL1
\usepackage{fullpage}
\fi

\newcommand{\startold}{\bgroup\color{orange} }
\newcommand{\stopold}{\egroup}

\newcommand{\ZZ}{{\mathbb Z}} 
\newcommand{\FF}{{\mathbb F}} 
\newcommand{\modplus}{{\oplus}_{p}\xspace} 
\newcommand{\modtimes}{{\otimes}_{p}\xspace} 

\newcommand{\gcc}{{GCC}\xspace}
\newcommand{\icc}{{ICC}\xspace}
\newcommand{\AS}{{\it AVX2 server}\xspace}
\newcommand{\AAS}{{\it AVX-512 server}\xspace}

\begin{document}

\if\HAL0
\title{High performance SIMD modular arithmetic for polynomial evaluation}
\else
\title{High performance SIMD modular arithmetic\\for polynomial evaluation}
\fi

\if\HAL0

\author[1,2]{Pierre Fortin*}

\author[2]{Ambroise Fleury}

\author[2]{Fran\c{c}ois Lemaire}

\author[3]{Michael Monagan}

\authormark{FORTIN \textsc{et al}}

\address[1]{Sorbonne Universit\'e, CNRS, LIP6, 
Paris, France}

\address[2]{Universit\'e de Lille, CNRS, Centrale Lille, CRIStAL, Lille,       
France}
\address[3]{Department of Mathematics, Simon Fraser University, 
Burnaby, B.C.,\\ V5A 1S6, Canada}

\corres{*Pierre Fortin, Sorbonne Universit\'e,
LIP6, 4 place Jussieu, 75252 Paris CEDEX 05, France. 
\email{pierre.fortin@sorbonne-universite.fr}}

\else
\author[ab]{Pierre Fortin}

\author[b]{Ambroise Fleury}

\author[b]{Fran\c{c}ois Lemaire}

\author[c]{Michael Monagan}

\affil[a]{Sorbonne Universit\'e, CNRS, LIP6, 
Paris, France}

\affil[b]{Universit\'e de Lille, CNRS, Centrale Lille, CRIStAL, Lille,       
France}

\affil[c]{Department of Mathematics, Simon Fraser University,

Burnaby, B.C., V5A 1S6, Canada

    Emails: pierre.fortin@sorbonne-universite.fr,
    ambroise.fleury.etu@univ-lille.fr,
    francois.lemaire@univ-lille.fr,
    mmonagan@cecm.sfu.ca
}    
\fi 

\if\HAL1
\maketitle
\fi

\if\HAL1
\begin{abstract}
\else
\abstract[Abstract]{
\fi
Two essential problems in Computer Algebra, namely polynomial
factorization and polynomial greatest common divisor computation, can
be efficiently solved thanks to multiple polynomial evaluations in two
variables using modular arithmetic. In this article, we focus on the
efficient computation of such polynomial evaluations on one single CPU
core. We first show how to leverage SIMD computing for modular
arithmetic on AVX2 and AVX-512 units, using both intrinsics and OpenMP
compiler directives. Then we manage to increase the  operational
intensity and to exploit instruction-level parallelism in order to
increase the compute efficiency of these polynomial evaluations.
All this results in the end to performance gains up to
about 5x on AVX2 and 10x on AVX-512.
\if\HAL1
\end{abstract}
\else
}
\fi

\if\HAL1
\def\keywords{\vspace{.5em}
{\textit{Keywords}:\,\relax%
}}
\def\endkeywords{\par}
\fi


\if\HAL0
\maketitle
\fi

\section{Introduction}

Computer Algebra, also called symbolic computation, consists of
developing algorithms and data structures for manipulating
mathematical objects in an \emph{exact} way.  Multivariate
polynomials with rational coefficients are essential objects in
Computer Algebra, and they naturally arise in many applications
(Mechanics, Biology, \ldots), especially in non-linear problems. Among
the classical operations on multivariate polynomials (sum, product,
quotient, \ldots), two non trivial operations are essential: polynomial
factorization and polynomial gcd (greatest common divisor)
computation\cite{Knu97,GCJ92,GG03}.  Those operations are necessary 
for solving polynomial systems and simplifying quotients of multivariate polynomials.

Modern algorithms\cite{MT19,HM16} for factorization and gcd computation
rely on many polynomial evaluations, which dominate the overall
computation cost.
These polynomial evaluations have two main features.
First, these are partial evaluations in the sense that
not all variables are evaluated:
given a polynomial with $n$ variables, we evaluate $n-2$ variables
which results in a polynomial with $2$ variables.
Second, the variables are evaluated at integers modulo a prime $p$ thus
all integer arithmetic operations are performed modulo $p$. Computing
modulo a 64 bit prime $p$ makes it possible 
to use machine integers and native CPU
operations, instead of arbitrary-precision integers.
Since these partial modular polynomial evaluations are currently a performance
bottleneck for polynomial factorizations and gcd computations, 
we aim in this article to speed-up their computation on 
modern CPUs.

We focus here on one compute server since most symbolic computations
are usually performed on personal workstations. 
We distinguish three main techniques to obtain performance gain on current
CPU compute servers\cite{HP17}:
\begin{itemize}
\item increasing the compute efficiency. This can be achieved by 
increasing the operational intensity, i.e. the number of operations per byte 
of memory (DRAM) traffic, which helps exploit modern HPC
architectures whose off-chip memory bandwidth is often the most performance
constraining resource \cite{WWP09}.
The compute efficiency can also be increased
by better filling the pipelined floating-point
units with instruction-level parallelism;
\item exploiting data-level parallelism on vector (or SIMD - single
instruction, multiple data) units. Such parallelism is increasingly
important in the
overall CPU performance since the SIMD vector width has been constantly
increasing from 64 bits (MMX\cite{MMX} and 3DNow!\cite{3DNow})
to 128 bits (SSE\cite{SSE},
AltiVec\cite{AltiVec}), then to 256 bits
(AVX and AVX2\cite{AVX2}), and recently to 512 bits 
(AVX-512\cite{AVX512}).  
For 64-bit integers, such AVX-512 SIMD units can now offer a 8x speedup
with respect to a scalar computation.
But efficiently exploiting such parallelism requires ``regular''
algorithms where the memory accesses and the computations are similar
among the lanes of the SIMD vector unit.
Moreover, multiple programming paradigms can be used for SIMD programming:
intrinsics, compiler directives, automatic vectorization, to name a few. 
For key applications, it is important to determine which programming
paradigm can offer the highest programming level (ease of programming
and portability) 
without sacrificing (too much) performance. 
\item exploiting thread-level parallelism on the multiple cores
and on the multiple processors available in a compute server.
This can be achieved thanks to multi-process, multi-thread or multi-task
programming.
\end{itemize}

Thread-level parallelism has already been introduced for the partial
modular polynomial evaluation (see e.g. Hu and Monagan\cite{HM16}).
We will therefore focus here
on single-core optimizations, namely increasing the compute efficiency and
SIMD parallelism, and we present in this article the following contributions.

\begin{itemize}

\item
Multiple algorithms have been designed to efficiently compute
modular multiplications in a scalar mode.
We first justify why the floating-point (FP) based algorithm with
FMA (fused multiply-add) of van der Hoeven et al.\cite{HLQ16} 
is best suited for current HPC
architectures, especially regarding SIMD computing. 
We show that the optimized AVX version implementation of van der Hoeven et
al.\cite{HLQ16}  can safely be used in our polynomial evaluation, and
we then propose the first (to our knowledge) 
implementation of such modular multiplication algorithm on AVX-512,
as well as the corresponding FP-based modular addition. 
With respect to the reference polynomial evaluation implementation of
Monagan and coworkers\cite{MT19,HM16}, which relies on scalar
integer-based algorithms for 
modular operations, we detail the performance gains of our SIMD FP-based 
modular arithmetic for modular operation microbenchmarks and for
the polynomial evaluation.

\item We carefully compare intrinsics and
OpenMP compiler directives for programming SIMD modular arithmetic and for
their integration within the polynomial evaluation.
We detail the relevance, the advantages, the issues and the
performance results of both programming paradigms. 

\item We show how to significantly improve the performance of
the 
modular polynomial evaluation by increasing
the operational intensity via data reuse,
and by filling 
the pipelines of the floating-point units. 
This is achieved thanks to the introduction of multiple ``dependent'' and
``independent'' evaluations and loop unrolling. We also show that
close-to-optimal performance can be obtained without extra memory
requirements. 
\end{itemize}

In the rest of this article, we first introduce in
Sect.~\ref{s:pres_evalpoly} our partial polynomial evaluation.
Then we detail in Sect.~\ref{s:SIMD} our integration of SIMD computing first in
modular arithmetic, then in the 
polynomial evaluation. 
Finally, we show how we can increase the compute efficiency of
our polynomial evaluation in Sect.~\ref{s:IncCompEff}.

\section{Modular polynomial evaluation in two variables}
\label{s:pres_evalpoly}

\subsection{Presentation}

We are given a multivariate polynomial $f \in \ZZ[x_1,\dots,x_n]$ and
we want to evaluate the variables $x_3,\dots,x_n$ in $f$ at integers modulo a prime $p$.
Let $\FF_p$ denotes the finite field of integers modulo a prime $p$.
The prime $p$ is chosen so that all integer arithmetic in $\FF_p$ 
can be done with the hardware arithmetic logic units 
of the underlying machine. 
For example, with a machine which supports a 64 bit by 64 bit integer multiply,
the application would use $p<2^{64}$.
We will pick non-zero integers $\beta_3,\beta_4,\dots,\beta_n$ uniformly at random from
$\FF_p$ and compute a sequence of $T$ evaluations
\begin{align*}
  b_t(x_1,x_2) =  f(x_1,x_2,\beta_3^t,\beta_4^t,\dots,\beta_n^t) ~~{\rm for}~~ 1 \le t \le T.
\end{align*}
The values $b_t$ are polynomials in $\FF_p[x_1,x_2]$.
We call them bivariate images of $f$.
For convenience we will use the notation
\begin{align*}
\beta   = (\beta_3,\beta_4,\dots,\beta_n) \in \FF_p^{n-2}, \textrm{\quad and \quad}
\beta^t = (\beta_3^t,\beta_4^t,\dots,\beta_n^t) \in \FF_p^{n-2}, 
\end{align*}
so that we may write $b_t(x_1,x_2) = f(x_1,x_2,\beta^t)$.

Before presenting two application examples of such computation, we
first emphasize that we evaluate here at powers of $\beta$, not at
$T$ different random points in $\FF_p^{n-2}$, since this enables one to save
computations. Indeed, when Zippel\cite{Zip79} introduced the first sparse
polynomial interpolation algorithm and
used his algorithm to compute a gcd of two polynomials $f$ and $h$ in
$\ZZ[x_1,\dots,x_n]$, he
used random evaluation points.
To interpolate a polynomial with $s$ terms his method solves one or more $s \times s$ linear
systems modulo $p$.  Using Gaussian elimination this does $O(s^3)$ arithmetic operations
in $\FF_p$ and requires space for $O(s^2)$ elements of $\FF_p$.
Then  Zippel\cite{Zip90} showed that if powers $\beta^t$ are used for the
evaluation points,
the linear systems are transposed Vandermonde systems which can be solved using
only $O(s^2)$ operations in $\FF_p$ and $O(s)$ space.
In Sect.~\ref{s:matrix_method}, we will see that 
using powers of $\beta$ also reduces the polynomial evaluation cost.

Second, we also emphasize that we evaluate at $\beta^t$ for $1 \le t \le T$,
and not for $0 \le t < T$, since
the evaluation point $\beta^0 = (1,1,\dots,1)$
may not be usable.
For example, consider the following gcd problem in $\ZZ[x_1,x_2,x_3,x_4]$.  Let
\begin{align*}
  f = g c \textrm{ ~and~} h = g c + g d
  \textrm{ ~~where~~} c = x_1 x_2 + x_3 x_4, d = x_3 - x_4
  \textrm{ ~and~} g \in \ZZ[x_1,x_2,x_3,x_4].
\end{align*}
Since $\gcd(c,d)=1$ we have
\begin{align*}
  \gcd(f,h) = \gcd( gc, gc + gd ) = \gcd( gc, gd ) = g \gcd(c,d) = g.
\end{align*}
But suppose we use $\beta^0 = (1,1)$.
Then since $d(x_1,x_2,1,1)=0$ 
we have
\begin{align*}
\gcd(f(x_1,x_2,\beta^0),h(x_1,x_2,\beta^0)) = g(x_1,x_2,1,1) c(x_1,x_2,1,1) = g(x_1,x_2,1,1) (x_1 x_2 + 1).
\end{align*}
We cannot interpolate $g$ using this image.  We say $(1,1)$ is an unlucky evaluation point.
Such 
unlucky evaluation points are avoided with high probability by picking
$\beta$ at random from $\FF_p^{n-2}$ 
and evaluating at $\beta^t$ for $t$ starting at $1$.

\subsection{Application examples}
\label{s:examples}

Such bivariate images of $f$ are 
needed in modern algorithms of
Computer Algebra for factoring polynomials with integer coefficients
and for computing gcd 
of polynomials with integer coefficients.
Two examples are presented below.

\subsubsection{Polynomial factorization}

Regarding polynomial factorization\cite{Knu97,GCJ92}, 
Monagan and Tuncer\cite{MT18,MT19} reduce factorization
of a multivariate polynomial
$f$ in $\ZZ[x_1,x_2,\dots,x_n]$ to (i) evaluating $f(x_1,x_2,\beta^t)$ for $1 \le t \le T$,
(ii) doing a computation with the bivariate images, and (iii) recovering the
factors of $f$ using sparse interpolation techniques\cite{Zip79,Zip90,BT88}.
See Roche\cite{Roche18} for a recent discussion on sparse polynomial
interpolation methods and an extensive bibliography.
If $f= \prod_{i=1}^{r} f_i$ is the factorization of $f$ over $\ZZ$ then usually the factors $f_i$
have a lot fewer terms than their product $f$.  Furthermore, because the method
interpolates the coefficients of the factors $f_i$ from bivariate images in $x_1$ and $x_2$,
the largest coefficient is likely to have a lot fewer terms than the factor $f_i$.
Because of this, the evaluation step dominates the cost.
The interpolation step, though more complicated,
is cheaper because the coefficients of the factors $f_i$ being interpolated have far fewer terms
than $f$ which is being evaluated.

We wish to give an example to illustrate the numbers involved.
We consider the factorization of
determinants of symmetric Toeplitz matrices\cite{MT19}.
The $m$'th symmetric Toeplitz matrix $T_m$ is an $m \times m$ matrix
in $m$ variables $x_0,x_1,\dots,x_{m-1}$ where
$T_{ij} = x_{|i-j|}$.
For example
\begin{align*}
T_4 = \left [ \begin{array}{cccc} x_0 & x_1 & x_2 & x_3 \\ x_1 & x_0 & x_1 & x_2 
                           \\ x_2 & x_1 & x_0 & x_1 \\ x_3 & x_2 & x_1 & x_0 \end{array} \right ].
\end{align*}
The problem is to factor the polynomial $\det T_m$.  For $m=14$, $\det T_m$
has $s=5,165,957$ terms.  It factors into two 
factors each with
$34,937$ terms.  The largest coefficient has
$u=9,705$ terms.
Thus $u$, the size of the interpolation problem, is 532 times smaller
than $s$, the size of the evaluation problem.

\subsubsection{Polynomial gcd}

Given two polynomials
$f,h \in \ZZ[x_1,x_2, \dots,x_n]$, to compute $g = \gcd(f,h)$, the parallel algorithm of 
Hu and Monagan\cite{HM16,HM19} works by computing bivariate images of $g$ modulo a prime $p$, 
namely,
\begin{align*}
g_t(x_1,x_2) = \gcd( f(x_1,x_2,\beta^t), h(x_1,x_2,\beta^t)) ~~\textrm{ for}~~ 1 \le t \le T.
\end{align*}
It then uses sparse interpolation techniques to interpolate $g$ from the images $g_t(x_1,x_2)$.
Since $g$ is a factor of $f$ and $h$, the number of terms in $g$ is typically
much fewer than the number in $f$ and $h$.
Let us use the notation $\#f$ for the number of terms of a polynomial $f$.
So for the gcd problem, typically $\#g \ll \max(\#f,\#h)$.
Hu and Monagan\cite{HM19} present a
``benchmark'' problem where $\#f=10^6$, $\#h=10^6$, and $\#g=10^4$.  If one interpolates $g$ from
univariate images then the largest coefficient of $g$ in $x_1$ has
$1108$ terms.
If instead, as the authors recommend, one interpolates $g$ from bivariate images,
then the largest coefficient of $g$ in $x_1$ and $x_2$ has only
$u=122$ terms.
So for this problem,
$u$ is almost 10,000 times smaller than $\max(\#f,\#h)$, 
the size of the evaluation problem.
Again, because of this, the authors found that the evaluations of the input
polynomials $f$ and $h$ completely dominate the cost of polynomial gcd computations.

\medskip 
Thus two very central problems in Computer Algebra, namely, polynomial
factorization
and polynomial gcd computation are usually dominated by evaluations when
there are many variables.

\subsection{The matrix method}
\label{s:matrix_method}

Let $p$ be a prime and let $f \in \FF_p[x_1,x_2,\dots,x_n]$.
We may write $f$ as
\begin{align}
f = \sum_{i=1}^{s} a_i x_1^{d_i} x_2^{e_i} M_i(x_3,\dots,x_n)
\label{e:def_f}
\end{align}
where $a_i \in \FF_p$ are non-zero, $d_i$ and $e_i$ are non-negative integers
and $M_i$ is a monomial in $x_3,\dots,x_n$.  For $\beta = (\beta_3,\beta_4,\dots,\beta_n) \in \FF_p^{n-2}$,
we want to compute $T$ partial evaluations 
\begin{align*}
b_t(x_1,x_2) = f(x_1,x_2,\beta^t) = \sum_{i=1}^{s} a_i x_1^{d_i} x_2^{e_i} M_i(\beta^t) {\rm ~~for~} 1\le t \le T.
\end{align*}
If we let $m_i = M_i(\beta_3,\dots,\beta_n) \in \FF_p$ and $M_i(x_3,\dots,x_n) = \prod_{k=3}^{n} x_k^{d_{ik}}$ then we observe that
\begin{align*}
M_i(\beta^t) = \prod_{k=3}^{n} (\beta_k^t)^{d_{ik}} = \prod_{k=3} (\beta_k^{d_{ik}})^t = M_i(\beta)^t = m_i^t.
\end{align*}
Thus
\begin{align*}
b_t(x_1,x_2) = f(x_1,x_2,\beta^t) = \sum_{i=1}^{s} a_i x_1^{d_i} x_2^{e_i} m_i^t.
\end{align*}
Now we can present the ``matrix method''\cite{HM16} which
relies on the powers of $\beta$ to efficiently compute the $T$
bivariate images. 
First we compute the monomial evaluation $m_i$ by evaluating $M_i(\beta_3,\dots,\beta_n)$.
To do this, let $d_k = \deg(f,x_k)$ and let $d = \max_{k=3}^{n} d_k$.
We pre-compute tables of powers 
\begin{align*}
  \left [ \beta_k^i {\rm ~for~} 0 \le i \le d_k \right ] ~~{\rm for}~ 3 \le k \le n.
\end{align*}
This takes at most $(n-2)(d-1)$  multiplications.
Then, for $1 \leq i \leq s$ we compute $m_i = M_i(\beta)$, using $n-3$
multiplications for each $m_i$ thanks to the tables of powers, 
and thus using 
$(n-3)s$ multiplications in total.
Therefore we can compute the $m_i$ with $O(nd+ns)$ multiplications.
Computing
\begin{align*}
b_1(x_1,x_2) = 
f(x_1,x_2,\beta) = \sum_{i=1}^{s} a_i m_i x_1^{d_i} x_2^{e_i}
\end{align*}
needs 1 multiplication for each $
a_i m_i \in \FF_p$, hence
$s$ multiplications in total. 
We can compute the next evaluation
\begin{align*}
b_2(x_1,x_2) = f(x_1,x_2,\beta^2) = \sum_{i=1}^{s} a_i m_i^2 x_1^{d_i} x_2^{e_i}
\end{align*}
using another $s$ multiplications if we save 
$a_i m_i \in \FF_p$ for $1 \le i \le s$ and
multiply them by $m_i$.  This leads to an algorithm that computes the $T$ evaluations
using $O(nd+ns)$ multiplications to compute the $m_i$, plus
a further $s T$ multiplications to compute the $a_i m_i^t$ for $1 \le t \le T,~
1 \le i \le s$, hence $O(nd+ns+sT)$ multiplications in total.
With random points instead of powers of $\beta$,
the $T$ evaluations would have required
a larger operation count of $O(nT(d+s))$ multiplications in total\cite{HM16}.

One way to see the matrix method is to think of evaluating $f$ at $\beta^t$ as the following $t \times s$ matrix-vector multiplication.
\begin{align*}
\left [ \begin{array}{cccc}  m_1  & m_2 & \dots & m_s \\
                            m_1^2 & m_2^2 & \dots & m_s^2 \\
                            \vdots & \vdots &\vdots & \vdots \\
                            m_1^T  & m_2^T  & \dots & m_s^T  \\
         \end{array} \right ] 
\left [ \begin{array}{c}
   a_1 x_1^{d_1} x_2^{e_1} \\ a_2 x_1^{d_2} x_2^{e_2} \\ \vdots \\ a_s x_1^{d_s} x_2^{e_s} 
 \end{array} \right ]
 = \left [ \begin{array}{c} b_1 \\ b_2 \\ \vdots \\ b_T \end{array} \right ]
\end{align*}
In practice, the complete matrix is not explicitly built and
we take advantage of the connection between successive rows of the matrix. 
Let $a = [ a_1, a_2, \dots, a_s]$, $m = [m_1, m_2, \dots,  m_s]$ and
$X = [ x_1^{d_1} x_2^{e_1}, x_1^{d_2} x_2^{e_2}, \dots, x_1^{d_s} x_2^{e_s}]$.
Let $u \circ v$ denote the Hadamard product of two vectors $u,v \in \FF_p^s$, that is
$ u \circ v = [u_1 v_1, u_2 v_2, \dots, u_s v_s] \in \FF_p^s. $
Then, viewing $b_1(x_1,x_2)$ and $b_2(x_1,x_2)$ as vectors
of terms, we have
\begin{align*}
b_1(x_1,x_2) = (a \circ m) \circ X \textrm{ ~~~~and~~~~} b_2(x_1,x_2) = ((a \circ m) \circ m) \circ X.
\end{align*}

Finally, for a given $t$ we will have to compute the sum $c_{i,t}$ 
of $a_i m_i^t$ for all $a_i m_i^t x_1^{d_i} x_2^{e_i}$
sharing the same $d_i$ and $e_i$ values. This sum $c_{i,t}$ is indeed the coefficient
of $x_1^{d_i} x_2^{e_i}$ in $b_t(x_1,x_2)$. These ``coefficient reductions'' are
required since in Eq. (\ref{e:def_f}), multiple $M_i(x_3,\dots,x_n)$
can potentially share the same $d_i$ and $e_i$ values.
If the monomials $x_1^{d_i} x_2^{e_i} M_i(x_3,\dots,x_n)$ in the input polynomial
$f$ are sorted
in lexicographical order with $x_1 > x_2 > x_3 > \dots > x_n$ then the monomials 
$x_1^{d_i} x_2^{e_i}$ will be sorted in $X$ which makes adding up $c_{i,t}$ 
coefficients of like monomials 
in $b_t(x_1,x_2)$ straightforward (with $O(s)$ additions for each evaluation).
Doing so, we compute
$f(x_1,x_2,\beta^t)$ for $1 \le t \le T$ with $O(nd+ns+sT)$ multiplications
and $O(sT)$ additions.

\begin{algorithm}[t]
    \begin{algorithmic}[1]
    \For{each evaluation $1 \le t \le T$} \label{eval_loop}
    \State $i \gets 1;$ $b_t \leftarrow 0$
    \While{$i \leq s$} 
        \State $c \gets 0$
        \State{$J \gets$ \#monomials with same $(d_i, e_i)$}
        \For{$i \le j < i+J $} \label{a:kernel:inner_loop}
          \State $a[j] \gets a[j] ~\modtimes~
          m[j]$ \label{a:kernel:op1} \Comment{Hadamard
          product\hspace*{\if\HAL0 8,5cm \else 4,5cm \fi}~}
          \State $c \gets c ~\modplus~
          a[j]$ \label{a:kernel:op2} \Comment{coefficient
          reduction\hspace*{\if\HAL0 8,5cm \else 4,5cm \fi}~} 
        \EndFor
        \State {\textbf{if} $c \neq 0$ \textbf{then} add $c x_1^{d_i}
        x_2^{e_i}$ to bivariate image $b_t$} \label{a:kernel:result}
        \State{$i \gets i+J$}
      \EndWhile
    \EndFor 
    \end{algorithmic}
    \caption{Compute kernel\label{a:kernel} of the matrix method
      for $T$ bivariate images of a polynomial
      $f \in \FF_p[x_1,x_2,\dots,x_n]$, 
      using notations of Eq. (\ref{e:def_f}).
      Inputs are the vector
      $m = [M_1(\beta), \dots, M_s(\beta)] \in \FF_p^s$ of monomial
      evaluations and the 
      coefficient vector $a = [a_1, \dots, a_s] \in \FF_p^s$.} 
\end{algorithm}

The resulting algorithm for the compute kernel of the matrix method
is detailed in Algorithm \ref{a:kernel}, where
we save the successive $a_i m_i^t$ values in the $a$ vector,
and where $\modplus$ and $\modtimes$ denote the arithmetic operators
modulo $p$ ($c = a \modtimes b $ denoting $c \equiv a \times b \pmod{p}$,
and $d = a \modplus b$ denoting $d \equiv a + b \pmod{p}$ with $(a,b,c,d)
\in \FF_p^4$).

In this article, we will use by default the following parameters when
measuring the time or performance of our partial modular polynomial
evaluation with 64-bit integers:
$s = 5 \times 10^5$ terms;
$n = 6$ variables, hence $4$ evaluated variables;
a maximum degree of $d=10$ in each variable;
and the number of evaluations $T$ chosen here as $10000$ to have a measurable computation time, but $T$ can be much lower in actual use.
These parameters have been chosen to be realistic and to lead to
stable and reproducible performance results.

\subsubsection{Multi-core parallel evaluation} 
\label{s:parallel_eval}
Hu and Monagan\cite{HM16} parallelized the matrix method for partial modular
polynomial evaluations on a multi-core architecture 
with $N$ cores by doing $N$ evaluations at a time. They first compute
$\Gamma = [m_1^N,m_2^N,\dots,m_s^N]$ using
exponentiations by squaring, requiring 
$O(s \log_2 N)$ multiplications.
Then they compute ${\Lambda}_k = a \circ [m_1^k,m_2^k,\dots,m_s^k]$ for $1 \le k \le N$
using $Ns$ multiplications.  Then, in parallel, the $k$'th core
successively computes $f(x_1,x_2,\beta^{k+N}) = {\Lambda}_k \circ \Gamma \circ X$;~
$f(x_1,x_2,\beta^{k+2N}) = {\Lambda}_k \circ \Gamma \circ \Gamma \circ X$;~ 
$f(x_1,x_2,\beta^{k+3N}) = {\Lambda}_k \circ \Gamma \circ \Gamma \circ \Gamma \circ X;~ \dots$ using Algorithm \ref{a:kernel} (with each ${\Lambda}_k$ as the $a$ vector,
and $\Gamma$ as the $m$ vector). 
This was implemented with multi-task programming in Cilk\cite{HM16}. 
This method significantly increases the space needed as
$N$ vectors $C_1,C_2,\dots,C_N$ of length $s$ are required where $s$ can be very large.
Monagan and Tuncer\cite{MT18} introduced an alternative parallelization 
strategy by using a 1D block decomposition of the $a$ and $m$ vectors for each
evaluation.

Finally, we mention the asymptotically fast method\cite{HL13} for computing
$f(x_1,x_2,\beta^t)$.  If $p$ is chosen of the form
$p-1=2^k q$ with $2^k>T$ so that an FFT of order $2^k$ can be done in the finite
field $\FF_p$, after computing the monomial evaluations $m=[m_1,m_2,\dots,m_s]$,
this method computes $f(x_1,x_2,\beta^t)$ for $1 \le t \le T$ 
in $O(s \, \log^2 T)$ multiplications.
Monagan and Wong\cite{MW17}  found that 
a serial implementation of this fast method first beat the matrix method
at $T=504$ but that it was much more difficult to parallelize than the matrix
method -- the fast method required $s \gg 10^6$ to deliver 
good parallel speedups.
Moreover, the simplicity and data locality of the matrix method makes
it very suitable for vectorization and other single-core optimizations targeted
in this article.

\section{SIMD modular arithmetic for partial polynomial evaluation}
\label{s:SIMD}

\subsection{Selection of the modular arithmetic algorithm}

Given a fixed\footnote{The value of $p$ is fixed in this
article, but is not a constant (from the programming point of view) in our
implementation. Our implementation will indeed be used for multiple $p$
values, which are unknown at compile time. The compiler cannot therefore
optimize
the code for a specific $p$ value.
} integer $p > 1$, we focus on the efficient computation of
$c = a \modtimes b $ and $d = a
\modplus b$\,.
We target the algorithm that will offer the best performance: 
such an algorithm must thus be efficient in scalar mode (i.e. non-SIMD),
while being also suitable for vectorization.
While $\modplus$ can be implemented with a compare instruction,
$\modtimes$ requires integer divisions
which are expensive operations 
on current processors\cite{Fog18}, and for which no SIMD integer
division instruction is available in SSE, AVX or AVX-512.
Hence, various
alternate algorithms have been designed in order to efficiently compute
$\modtimes$. 
We briefly recall 
the most important ones.

In order to compute $c = a \modtimes b $, 
one can first rely on floating-point arithmetic to compute $q = \lfloor \frac{a \times b}{p} \rfloor$ and then deduce $c = a \times b - q \times p$ 
(see for example Alverson\cite{Alv91}, Baker\cite{Bak92}).
This requires
conversions to/from floating-point numbers, and the number of bits of the floating-point number mantissa has to be twice as large as the number of bits of $p$
(to hold the product).
In order to avoid the conversions between floating point numbers and integers,
the floating-point reciprocal $p^{-1}$ 
can be rescaled and truncated into an integer.
The quotient $q$ is hence approximated, and some adjustments enable one to
obtain the remainder $c$.
This integer-based method is known as the {\it Barrett's
product}\cite{Bar87,GM94}.  
Another integer-based approach relies on Montgomery's reduction\cite{Mon85}:
a comparison between the two methods has been done for example by
van der Hoeven et al.\cite{HLQ16}.
An improved version of Barrett's product
with integer only operations has been proposed
by M\"oller and Granlund\cite{MG11}: Monagan and coworkers\cite{HM16,MT18}
use an implementation (written by Roman Pearce) of this
latter\cite{MG11} method (with $p^{-1}$ precomputed) 
 in their original code for polynomial evaluation with 64-bit
 integers.
 This offers a 11x performance gain\cite{GM15} for $\modtimes$ with
 respect to one integer division.
   However, this implementation relies on 128-bit intermediate results: 
for SIMD processing on 64-bit elements, this implies that only
half of the SIMD lanes will be used, hence leading to twice lower SIMD
speedups. One can replace these 128-bit variables with two 64-bit variables
(hence using only one 64-bit lane per operation), but this requires extra
arithmetic and
bit shifting.
Moreover, to our knowledge
there is no SSE/AVX2/AVX-512 intrinsic which performs
multiplications on 64-bit integers and provides either the 128-bit results
(similarly to the \texttt{\_mm\{,256,512\}\_mul\_epu32} intrinsics on
32-bit integers) or their upper and lower 64-bit parts.
This greatly complicates the SIMD programming of 
the M\"oller and Granlund\cite{MG11} algorithm for 64-bit integers.

\begin{algorithm}[t]
    \caption{\label{a:modtimes}
    Modular multiplication of 64-bit integers $x$ and $y$
    with a 50-bit prime $p$. $x$ and $y$ are considered to be already reduced
    modulo $p$, and converted to \texttt{double} along with $p$
    prior to the beginning of the
    algorithm.
    $u$ stores: \texttt{1/(double) p} }
    \begin{algorithmic}[1]
    \State double~ $ h \gets x * y$\,; \label{a:modtimes:h}
    \State double~ $ \ell \gets$ fma($x$, $y$, $-h$)\,; \label{a:modtimes:l}
    \State double~ $ b \gets h * u$\,;
    \State double~ $ c \gets$ floor($b$)\,; \Comment{$c$ is the
    quotient $\pm 1$\hspace*{\if\HAL0 8,5cm \else 4,5cm \fi}~} 
    \State double~ $ d \gets$ fma($-c$, $p$, $h$)\,;
    \State double~ $ g \gets d+\ell$\,; \label{a:modtimes:g} \Comment{$g$ is the remainder $\pm p$\hspace*{\if\HAL0 8,5cm \else 4,5cm \fi}~} 
    \State {\textbf{if} $g \geq p$ \textbf{then} return $g-p$\,;} \label{a:modtimes:test1} 
    \State {\textbf{if} $g < 0.0$ \textbf{then} return $g+p$\,;} \label{a:modtimes:test2} 
    \State {return $g$\,;}
    \end{algorithmic}
\end{algorithm}

  \begin{algorithm}[t]
  \caption{\label{a:modplus}
    Modular addition of 64-bit integers $x$ and $y$
    with a 50-bit prime $p$. $x$ and $y$ are considered to be already reduced
    modulo $p$, and converted to \texttt{double} along with $p$
    prior to the beginning of the
    algorithm.}
    \begin{algorithmic}[1]
    \State double~ $ s \gets x + y$\,;
    \State {return~ $s \geq p$ ? $s-p$ : $s$ \,;} \label{a:modplus:test}      
  \end{algorithmic}
\end{algorithm}

\smallskip

We therefore focus in this article
on the use of floating-point (FP) FMA (fused multiply-add)
instructions for floating-point based modular arithmetic. 
Since the FMA instruction performs two operations ($a * b + c$)
with one single final rounding, it can indeed be used to design
a fast {\it error-free transformation} of the product of two floating-point
numbers\cite{ORO05}. Such error-free transformation computes the accurate
floating-point result of the product.
As described and proved by van der Hoeven et al.\cite{HLQ16},
this makes it possible 
to design a modular multiplication with double-precision floating-point
numbers, provided that $p$ has at most 50 bits: see Algorithm \ref{a:modtimes}. 
Intuitively, an error-free transformation
(Lines \ref{a:modtimes:h} and \ref{a:modtimes:l} in Algorithm \ref{a:modtimes})
enables one to compute in twice working precision\cite{ORO05},
and hence to precisely handle the multiplication result 
before reduction modulo $p$.
More precisely, $\ell$ stores the rounding error of the product $x * y$
(i.e. $h+\ell$ exactly equals $x * y$).
The approximate real quotient $(x*y)/p$ is then computed in $b$ using the
pre-computed $u = 1/(double) p$, and rounded to an (approximate)
integer quotient $c$.
A first approximate remainder $d$ is computed using $c * p$,
and added to $\ell$ in $g$ in order
to take into account the initial rounding error.
$g$ is finally corrected so that we  exactly have: $g  \equiv x \times
y \pmod{p}$.
We emphasize that all this is achieved with 64-bit floating-point
numbers only: no larger variables are required and
we can thus benefit from the full SIMD speedup (up to 8x on
AVX-512).
The corresponding FP-based modular addition algorithm is presented in
Algorithm \ref{a:modplus}.

We also emphasize that the limit on the size of $p$ (at most 50 bits) is not
problematic regarding our targeted applications (presented in
Sect.~\ref{s:examples}).
Indeed, let $f(x_1,x_2,\dots,x_n) = \sum_{i=1}^s a_i M_i(x_1,x_2,\dots,x_n)$ be
a polynomial we wish to interpolate.
Ben-Or/Tiwari 
\cite{BT88} 
and Zippel
\cite{Zip90}
both pick
$\beta \in \FF_p^n$ at random and interpolate $f$ from
$f(\beta),f(\beta^2),f(\beta^3),\dots.$ 
Both methods require the monomial evaluations to be distinct,
that is, $M_i(\beta) \ne M_j(\beta)$ for $1 \le i < j \le s$.
For this to hold with reasonable probability we require $p > 100 s^2$.
For a large value of $s$, say $10^4 < s < 10^6$, the requirement
$p>100 s^2$ means 32-bit primes are too small
but 50-bit primes are sufficient.

Finally, we stress 
that relying on FMAs is relevant regarding
HPC architectures. Current high-end
HPC-oriented Intel CPUs with AVX2 or AVX-512 indeed offer two FMA SIMD units.
HPC-oriented GPUs from NVIDIA or AMD (not studied in this article),
whose performance strongly
depends on SIMD computing, also fully support FMA instructions.

\subsection{SIMD programming paradigms}
\label{s:paradigms}

We plan to integrate the SIMD implementations of
the FMA-based $\modplus$ and $\modtimes$ modular operations  in
the polynomial evaluation algorithm (see Algorithm \ref{a:kernel}) on
AVX2 or AVX-512 CPUs.
For this purpose, there are multiple programming paradigms regarding
SIMD computing. 

A first possibility is to rely on intrinsics programming. Such
low-level programming enables the programmer to reach high
performance, but at a non-negligible development cost. 
This will be our primary programming paradigm, and we will detail
the corresponding implementations in Sect.~\ref{s:avx-512}.

A second possibility is to rely on the compiler to benefit from a
higher programming level. 
Compilers can detect parallel and vectorizable loops and automatically
vectorize these loops. But, as further detailed in
Sect.~\ref{s:SIMD4evalnext2},
such automatic vectorization will fail in our polynomial evaluation. 
Therefore, we consider here a third programming paradigm:
compiler directives for SIMD programming.
In C/C++ programming, these are pragmas  which enable the programmer
to indicate (and ensure) that a given loop is parallel: no dependency
analysis is then required by the compiler. 
Such compiler directives are
available in the Intel C/C++ Compiler \icc (\texttt{\#pragma simd}), and have
been standardized in the last versions of
OpenMP\footnote{\url{https://www.openmp.org/}}
(starting from OpenMP 4.0). 
We will rely here on OpenMP due to its sustainability, its wide usage
in HPC, and its availability in both \icc and \gcc
(the GNU Compiler Collection).
Such high level programming with OpenMP will enable us to avoid
writing intrinsics, to have one scalar C code for both AVX2 and AVX512,
and to avoid array padding or loop splitting when the iteration number
is not a multiple of the SIMD vector size. OpenMP directives will
also enable us to overcome the limits of the automatic vectorization
for our polynomial evaluation. 
However, the SIMD code generated by the compiler may differ from the
intrinsic code and hence lead to lower performance.

In the rest of Sect.~\ref{s:SIMD},
we will thus investigate these
two SIMD programming
paradigms: SIMD intrinsics and OpenMP SIMD directives.
Their performance results will also be detailed and compared.

\subsection{SIMD intrinsics and the AVX-512 version}
\label{s:avx-512}

\paragraph{Using AVX intrinsics:}

Van der Hoeven et al.\cite{HLQ16} have presented a SSE/AVX version
of Algorithm \ref{a:modtimes} to implement $\modtimes$. They use two SSE/AVX
\texttt{blendv\_pd} intrinsics to efficiently implement the two final tests
(Lines \ref{a:modtimes:test1}-\ref{a:modtimes:test2} in Algorithm
\ref{a:modtimes}), hence removing divergence in the SIMD computation.
This \texttt{blendv\_pd} intrinsic 
blend \texttt{double} elements from two vectors depending on
the most significant bit of elements from a third vector.
For floating-point elements this most significant bit corresponds
to the sign bit, which enables one to implement in SIMD without branching
the two final tests using comparisons to 0.0.
We will rely on this AVX version on AVX2 CPUs.
However, we recall that IEEE standard 754 for floating-point
arithmetic\cite{IEEE-Norm} includes signed zeros which may
lead to incorrect results regarding the use of  \texttt{blendv\_pd}.
Indeed if for example $g$ equals $-0.0$ at Line \ref{a:modtimes:test2} 
in Algorithm \ref{a:modtimes}, using the \texttt{blendv\_pd}
instruction directly on $g$ would return $p$, which
is incorrect for modulo $p$ arithmetic. 
We show 
below that $-0.0$ cannot appear in our
specific context: we can thus safely use this implementation with AVX
intrinsics.

Van der Hoeven et al. also present an AVX
$\modplus$ implementation (see function 3.9\cite{HLQ16}) of the scalar
FP-based $\modplus$ algorithm 
(presented in Algorithm \ref{a:modplus}). Similarly, as shown below, 
we can safely ignore signed zeros for the 
    the \texttt{blendv\_pd}
   instruction used in this AVX version.

 \paragraph{
 Regarding the issue with signed zeros and the AVX \texttt{blendv\_pd}
 intrinsic:} 

  For $\modtimes$, we consider the tests at Lines \ref{a:modtimes:test1}
  ($g \ge p$ here rewritten as $g-p \ge 0$) 
  and \ref{a:modtimes:test2} ($g > 0$) in Algorithm \ref{a:modtimes}. 
  We aim at showing here that no $-0.0$ value will occur in these
  two tests, so that one can safely use the 
   \texttt{blendv\_pd} intrinsics to implement  in AVX the conditional
   affectation resulting from these tests.
  For completeness, we recall beforehand the paragraphs 3 and 4 of
  \S6.3, {\it The sign bit}, from the IEEE 754
  standard\cite{IEEE-Norm}. 

      Paragraph 3: 
    {\it  When the sum of two operands with opposite signs (or the difference
  of two operands with like signs) is exactly zero, the sign of that
  sum (or difference) shall be +0 in all rounding-direction attributes
  except roundTowardNegative; under that attribute, the sign of an
  exact zero sum (or difference) shall be $-0$.  However, x + x = x $-$ ($-$x)
  retains the same sign as x even when x is zero.}

    Paragraph 4: 
  {\it When (a * b) + c is exactly zero, the sign of fusedMultiplyAdd(a, b,
  c) shall be determined by the rules above for a sum of
  operands. When the exact result of (a * b) + c is non-zero yet the
  result of fusedMultiplyAdd is zero because of rounding, the zero
  result takes the sign of the exact result.}
    \smallskip
    
  Now, let us first consider the case where $x$ or $y$ is zero (possibly
  both).  Then $h$ computed at Line \ref{a:modtimes:h} is either $+0.0$ or
  $-0.0$. According to the paragraph 4 quoted just above,
  the $\ell$ computed at Line \ref{a:modtimes:l} is obtained from the
  computation $h-h$ which cannot result in $-0.0$ thanks to
  paragraph 3 and since we rely on the default rounding mode (Round to
  nearest). Thus  $\ell=+0.0$.
  Using again paragraph 3,
  the $g$ computed at Line \ref{a:modtimes:g} by
  $g=d+\ell$ cannot be equal to $-0.0$ since $\ell=+0.0$. Therefore
  $g$ at Line \ref{a:modtimes:g} equals $+0.0$ (since the expected
  result of the modular product is zero here): thus no $-0.0$ is
  evaluated in the two tests at Lines \ref{a:modtimes:test1}
  and \ref{a:modtimes:test2}. 

  Second, consider the case were both $x$ and $y$ are nonzero.
  Since $p$ is a prime number, and since $x < p$ and $y < p$,
  the product $x \times y$ cannot
  be zero modulo $p$: this is due to the uniqueness of the prime
  factorization of $x \times y$. 
  Thus the function returns a nonzero value that is not a multiple of
  $p$. Consequently $g$ cannot hold $-0.0$, and neither can
  $g-p$ evaluate to $-0.0$. 

\smallskip

For $\modplus$, 
the \texttt{blendv\_pd} intrinsic evaluates the result of a subtraction 
by $p$ since the test $s \ge p$ at Line \ref{a:modplus:test} in
Algorithm \ref{a:modplus} is rewritten as $s-p \ge 0$
(see function 3.9\cite{HLQ16}). 
According to paragraph 3, this subtraction 
cannot lead to $-0.0$ since $p \neq -0.0$.

\begin{algorithm}[t]
  \begin{algorithmic}[1]
    \State \_\_m512d $\bar{h} \gets$  \_mm512\_mul\_pd($\bar{x}$, $\bar{y}$)\,;
    \State \_\_m512d $\bar{\ell} \gets$ \_mm512\_fmsub\_pd($\bar{x}$, $\bar{y}$, $\bar{h}$)\,;
    \State \_\_m512d
    $\bar{b} \gets$ \_mm512\_mul\_pd($\bar{h}$, $\bar{u}$)\,; 
    \State \_\_m512d $\bar{c} \gets$ \_mm512\_floor\_pd($\bar{b}$)\,;     
    \State \_\_m512d $\bar{d} \gets$ \_mm512\_fnmadd\_pd($\bar{c}$, $\bar{p}$, $\bar{h}$)\,; 
    \State \_\_m512d $\bar{g} \gets$ \_mm512\_add\_pd($\bar{d}$, $\bar{\ell}$)\,;     
    \State \_\_mmask8 $m \gets$ \_mm512\_cmplt\_pd\_mask($\bar{g}$, \_mm512\_setzero\_pd())\,;
    \State \_\_mmask8 $mm \gets$ \_mm512\_cmple\_pd\_mask($\bar{p}$, $\bar{g}$)\,;
    \State $\bar{g} \gets$ \_mm512\_mask\_add\_pd($\bar{g}$, $m$, $\bar{g}$, $\bar{p}$)\,;
    \State $\bar{g} \gets$ \_mm512\_mask\_sub\_pd($\bar{g}$, $mm$, $\bar{g}$, $\bar{p}$)\,;
    \State {return $\bar{g}$\,;}
    \end{algorithmic}
    \caption{\label{a:modtimes_AVX512}
      AVX-512 modular multiplication of AVX-512 vectors $\bar{x}$ and $\bar{y}$
      with a 50-bit prime $p$ replicated in the AVX-512 vector $\bar{p}$.
      Elements of $\bar{x}$ and $\bar{y}$ are considered
      to be already reduced modulo $p$, and converted (like $\bar{p}$)
      to \texttt{double}
      elements ($\bar{x}$, $\bar{y}$, $\bar{p}$ and $\bar{u}$ being \texttt{\_\_m512d} vectors)
      prior to the beginning of the algorithm.
      $\bar{u}$ stores replicates of: \texttt{1/(double) p} }
\end{algorithm}

\begin{algorithm}[t]
  \begin{algorithmic}[1]
    \State \_\_m512d $\bar{s} \gets$  \_mm512\_add\_pd($\bar{x}$, $\bar{y}$)\,;
    \State \_\_mmask8 $m \gets$ \_mm512\_cmple\_pd\_mask($\bar{p}$, $\bar{s}$)\,;
    \State {return \_mm512\_mask\_sub\_pd($\bar{s}$, $m$, $\bar{s}$, $\bar{p}$)\,;}
    \end{algorithmic}
    \caption{\label{a:modplus_AVX512}
      AVX-512 modular addition of AVX-512 vectors $\bar{x}$ and $\bar{y}$
      with a 50-bit prime $p$ replicated in the AVX-512 vector $\bar{p}$.
      Elements of $\bar{x}$ and $\bar{y}$ are considered
      to be already reduced modulo $p$, and converted (like $\bar{p}$)
      to \texttt{double}
      elements ($\bar{x}$, $\bar{y}$ and $\bar{p}$
      being \texttt{\_\_m512d}  vectors)
      prior to the beginning of the algorithm.}
\end{algorithm}

\paragraph{Using AVX-512 intrinsics:}
Regarding AVX-512, the \texttt{blendv\_pd} intrinsic is 
not available:
we therefore explicitly build 8-bit masks to conditionally
perform (without branching) the addition and the
subtraction at the end of the algorithm,
as presented in Algorithm \ref{a:modtimes_AVX512}.
Hence the SIMD divergence is efficiently handled within the AVX-512
arithmetic instructions. 
To our knowledge this is the first AVX-512 floating-point based modular
arithmetic. Orisaka et al.\cite{OAL18} have also accelerated modular
arithmetic with  AVX-512 but using Montgomery reduction and targeting
very large primes for cryptography.

Regarding $\modplus$, we also adapt the SSE/AVX implementation for
floating-point numbers presented by Van der Hoeven et al.\cite{HLQ16}  
 to AVX-512. As detailed in
Algorithm \ref{a:modplus_AVX512}, we use 1 addition, 1 comparison
and 1 masked subtraction in AVX-512 instead of 1 addition, 1 subtraction
and 1 \texttt{blendv\_pd} in AVX.

\subsection{Microbenchmarks}
\label{s:ubenchs}

\if\HAL0
\def\longueur{0.49}
\else
\def\longueur{0.48}
\fi 
\begin{figure}[t]

    \centering
    \subfigure[$\modtimes$ on \AS.]{%
    \includegraphics[width=\longueur\linewidth]{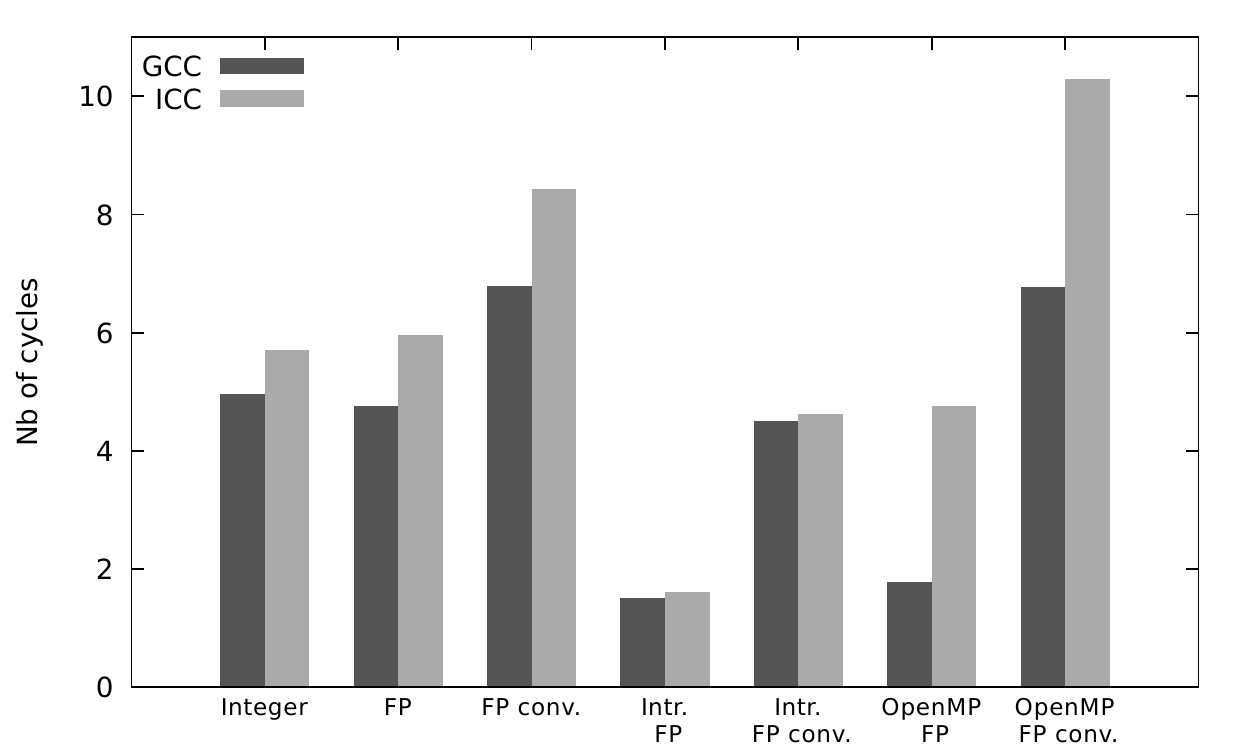}\label{f:microbenchs_times_AS}
    }
    \hfill
    \subfigure[$\modtimes$ on \AAS.]{%
    \includegraphics[width=\longueur\linewidth]{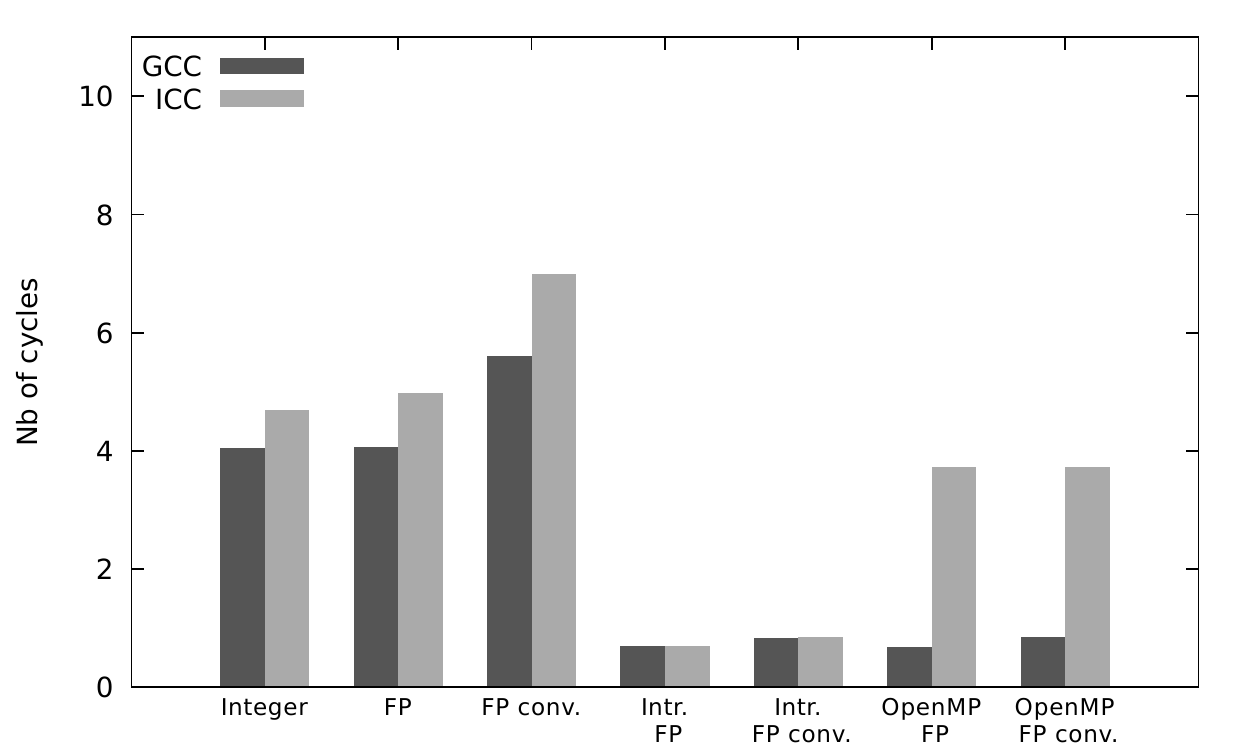}\label{f:microbenchs_times_AAS}
    }

    \centering
    \subfigure[$\modplus$ on \AS.]{%
    \includegraphics[width=\longueur\linewidth]{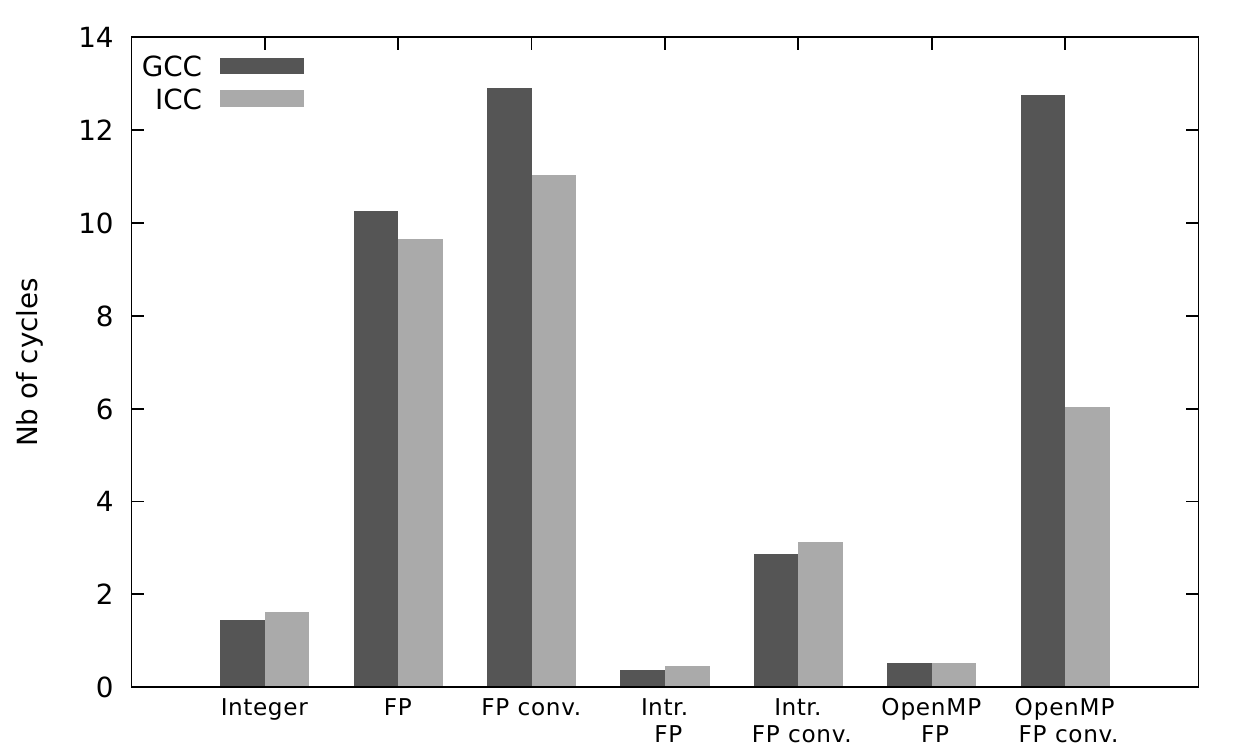}\label{f:microbenchs_plus_AS}
    }
    \hfill
    \subfigure[$\modplus$ on \AAS.]{%
    \includegraphics[width=\longueur\linewidth]{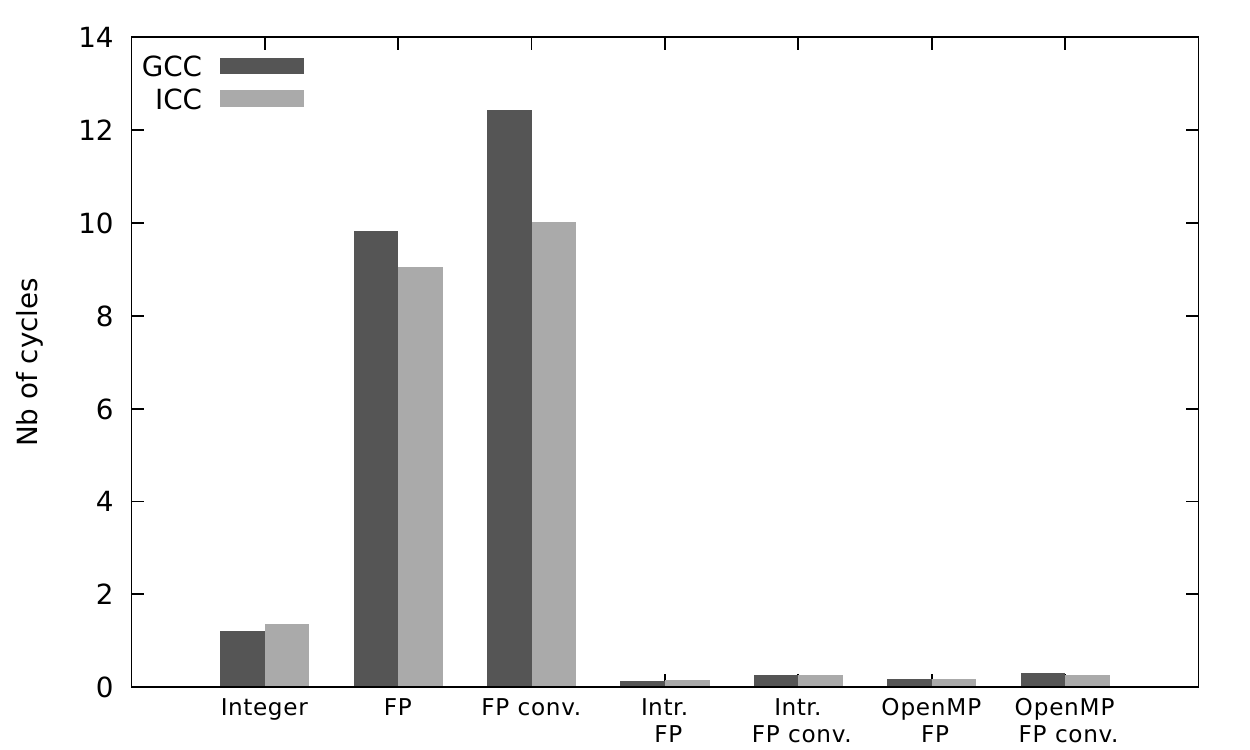}\label{f:microbenchs_plus_AAS}
    }

    \caption{Performance comparison of various implementations for
      $\modtimes$ and $\modplus$.
      \texttt{Integer} denotes the original integer-based
      implementation 
      as used by Monagan and coworkers for $\modtimes$\cite{HM16,MT18}  
      and $\modplus$\cite{LM15}.
      \texttt{FP} denotes the scalar floating-point based implementation without
      considering the conversions between integers and floating-points
      numbers, whereas \texttt{FP conv.} includes these conversions.
      \texttt{Intr. FP} (respectively \texttt{OpenMP FP})
      denotes the SIMD version of the
      floating-point based implementation using intrinsics (resp. using OpenMP).
      These performance results 
      have been obtained on element-wise operations over vectors
      of 2048 elements. 
    }\label{f:microbenchs}
    
\end{figure}

\begin{table}[t]
\caption{Test platforms.\label{t:platforms}} 
\if\HAL1
{\small
\begin{center}
\fi
  \begin{tabular}{l|l|l}
\hline
     Servers & Name: & Hardware features:\\
\if\HAL0
& \AS  & 2$\times$ Intel Xeon CPU E5-2695 v4 CPUs: 2$\times$18  2-way SMT cores - 2.10 GHz (base) / 3.30 GHz (turbo) - AVX2 \\         
\else
& \AS  & 2$\times$ Intel Xeon CPU E5-2695 v4 CPUs: 2$\times$18  2-way SMT \\
& & $\quad$  cores - 2.10 GHz (base) / 3.30 GHz (turbo) - AVX2 \\         
\fi
\if\HAL0
     & \AAS  & 2$\times$ Intel Xeon Gold 6152 CPUs: $\qquad~~$ 2$\times$22  2-way SMT cores - 2.10 GHz (base) / 3.70 GHz (turbo) - AVX512 \\
\else
     & \AAS  & 2$\times$ Intel Xeon Gold 6152 CPUs: $\qquad~~$ 2$\times$22  2-way SMT \\
     & & $\quad$ cores - 2.10 GHz (base) / 3.70 GHz (turbo) - AVX512 \\
\fi
\hline
\hline
     Compilers & Name: & Performance-related options:\\
     & \gcc 8.2.0 & \texttt{-O3 -mfma -fno-trapping-math -march=native -mtune=native} \\
     & \icc 19.0.3.199 & \texttt{-O3 -fma -xhost}\\
\hline
\end{tabular}
\if\HAL1
\end{center}
}
\fi 
\end{table}

We start with microbenchmarks,  presented in 
Fig. \ref{f:microbenchs}, of the two modular arithmetic
operations  $\modtimes$ and $\modplus$ .
Like all following performance tests, these microbenchmarks
have been performed on the two compute
servers \AS and \AAS presented in Table \ref{t:platforms}.
We first notice that,
regarding the original implementation of the modular multiplication (integer based, by R. Pearce), our microbenchmark results are consistent with the
  6.016 cycles obtained with \gcc by Monagan and coworkers\cite{GM15}
  on older CPUs.

  Regarding $\modtimes$ (see Figs. \ref{f:microbenchs_times_AS} and
  \ref{f:microbenchs_times_AAS}),
  and with respect to the original integer based implementation, 
  the scalar floating-point (FP) based implementation
  offers lower performance
  when
  including back and forth conversions between integers and floating-points
  numbers, and similar performance when not considering these conversions. 
  The SIMD FP based implementation offer (with intrinsics, and
  without considering the conversions)
  a 3.2x
  (resp. 3.7x) 
  speedup with \gcc (resp. \icc)
  over the scalar FP-based implementation on \AS.
  On \AAS, the speedup is 5.9x 
  (resp. 7.2x)
  speedup with \gcc (resp. \icc). 
  This shows that the performance gain of our new AVX-512 $\modtimes$
  implementations is indeed
  twice greater than the AVX2 one.

 Regarding $\modplus$ (see Figs. \ref{f:microbenchs_plus_AS} and
  \ref{f:microbenchs_plus_AAS}), the scalar FP-based implementation
  leads to much greater
  cycle numbers than the integer-based one:
  this is due to the branching of the compare instruction
  required in the FP implementation (see Algorithm \ref{a:modplus}).
  The comparison in the integer-based $\modplus$ implementation can
  indeed be replaced by shifting\cite{LM15,HLQ16}, which avoids the
  branching performance impact on the pipeline filling.
  Thanks to the use of the AVX2 \texttt{blendv\_pd} intrinsic
  and of AVX-512 masks, there is also no branching in the SIMD
  FP-based implementations (with intrinsics) which implies 
  a strong performance gain (around one order of magnitude),
  in addition to the SIMD speedup.
  With respect to the scalar integer-based one,
  the SIMD speedups of the FP-based implementations
  with intrinsics (without considering the
  conversions) are 4.0x
  (resp. 3.6x)
  with \gcc (resp. \icc) on \AS, and
  8.7x 
  (resp. 8.5x)
  with \gcc (resp. \icc) on \AAS. 
  This shows that our AVX-512 $\modplus$ implementation is
  twice faster than the AVX2 one of Van der Hoeven et al.\cite{HLQ16}.

 When considering the conversions, the overhead of these conversions
  can annihilate the SIMD performance gain on \AS. This is due to the lack
  of AVX2 conversion instruction between 64-bit integers and 64-bit
  floating-point numbers: the conversions are thus performed in scalar mode
  which has a strong performance impact.
  In comparison, such a SIMD instruction is
  available in AVX-512\footnote{More precisely, the 64-bit conversions
    belong to the AVX-512DQ instruction set which is available on our Intel
    Xeon Gold 6152 CPUs, but not on the prior Intel
    Knights Landing (Xeon Phi) processors.
    }, where conversions 
  can be performed in SIMD mode. Their overheard is therefore much lower 
  on \AAS. 

\smallskip

 Finally, we also consider using OpenMP to vectorize the code. 
  The first issue lies in having the compiler generate SIMD FMA instructions
  from the \texttt{fma()} function call in the C+OpenMP code for $\modtimes$.
  This is
  effective with \gcc thanks to the \texttt{-fno-trapping-math} option
  which allows us to assume that floating-point operations cannot generate
  traps, such as division by zero, overflow, underflow, inexact result and
  invalid operation. Unfortunately, using all possible floating-point
  model variations (\texttt{-fp-model} options) did not enable us to
  generate SIMD FMA instructions with \icc. The \icc OpenMP code hence
  relies on scalar FMA instructions, which explains its important
  performance overhead over the \gcc OpenMP code for $\modtimes$. 
  Secondly, on \AAS we had to force the AVX-512 vectorization using 
  \verb?-qopt-zmm-usage=high? with \icc and
  \verb?-mprefer-vector-width=512? with \gcc, otherwise only AVX2
  instructions are generated.

  As far as $\modtimes$ is concerned, the OpenMP code (with \gcc) has the same
  performance than the SIMD code written in intrinsics on \AAS, but is
  slower by 18\% on \AS. This is because of one additional compare
  instruction 
   added by the compiler before each \texttt{blendv\_pd} instruction.
   This comparison to zero (either $g > 0$, or $g-p \ge 0$) 
   is here to prevent any issue with IEEE
   754 signed zeros
    and the \texttt{blendv\_pd} instruction.
    The compiler is indeed unaware of our specific context which
   enables us not to consider $-0.0$, as shown in Sect.~\ref{s:avx-512}.
   
   For $\modplus$, the compiler similarly adds 
   one unnecessary compare instruction which results in a 45\%
   performance penalty on \AS for the OpenMP code (with \gcc). 
   However no branching instruction is generated in the SIMD code for
   $\modplus$ with OpenMP, which makes this OpenMP still rather
   efficient with respect to the scalar integer-based implementation:
   2.8x faster  
   on \AS, and
   7.4x 
   on \AAS (with \gcc).

\subsection{Integration in polynomial evaluation}
\label{s:SIMD4evalnext2}

\begin{algorithm}[t]
    \begin{algorithmic}[1]
    \State In-place conversions for vectors $a$ and $m$ (64-bit integers $\rightarrow$ doubles)   
    \For{each evaluation $1 \le t \le T$} \label{a:kernel_simd:eval_loop}
    \State $i \gets 1;$ $b_t \leftarrow 0$
    \While{$i \leq s$} 
        \State $\bar{c} \gets 0.0$
        \State{$J \gets$ \#monomials with same $(d_i, e_i)$}
        \For{$i \le j < i+J $ with step ${\cal V}$} \label{a:kernel_simd:inner_loop}
          \State $ \bar{a} \gets a[j~..~j+{\cal V}-1]$ \Comment{SIMD load\hspace*{\if\HAL0 6,8cm \else 3,5cm \fi}~}   
          \State $ \bar{m} \gets m[j~..~j+{\cal V}-1]$ \Comment{SIMD load\hspace*{\if\HAL0 6,8cm \else 3,5cm \fi}~}  
          \State $ \bar{a} \gets \bar{a} ~\modtimes~ \bar{m}$ \label{a:kernel_simd:op1} \Comment{SIMD Hadamard product\hspace*{\if\HAL0 6,8cm \else 3,5cm \fi}~} 
          \State $ \bar{c} \gets \bar{c} ~\modplus~ \bar{a}$ \label{a:kernel_simd:op2} \Comment{SIMD coefficient reduction\hspace*{\if\HAL0 6,8cm \else 3,5cm \fi}~} 
          \State $ a[j~..~j+{\cal V}-1] \gets \bar{a}$ \Comment{SIMD store\hspace*{\if\HAL0 6,8cm \else 3,5cm \fi}~} 
          \EndFor
          \State $c \gets reduce(\bar{c}, ~\modplus)$ \label{a:kernel_simd:final_reduction} \Comment{$\bar{c}$ final reduction\hspace*{\if\HAL0 6,8cm \else 3,5cm \fi}~} 
          \State {\textbf{if} $c \neq 0.0$ \textbf{then} convert $c$ to
          64-bit integer and add $c x_1^{d_i}
        x_2^{e_i}$ to bivariate image $b_t$} \label{a:kernel_simd:result}
        \State{$i \gets i+J$}
      \EndWhile
    \EndFor 
    \end{algorithmic}
    \caption{SIMD compute kernel\label{a:kernel_simd} of the matrix
    method (see Algorithm \ref{a:kernel} for notations and
    inputs). 
      $\bar{x}$ denotes the SIMD vector corresponding to variable
      $x$. 
      ${\cal V}$ is the size of the SIMD vector. 
      }     
\end{algorithm}

\begin{figure}[t]

    \centering
    \subfigure[On \AS.]{%
    \includegraphics[width=\longueur\linewidth]{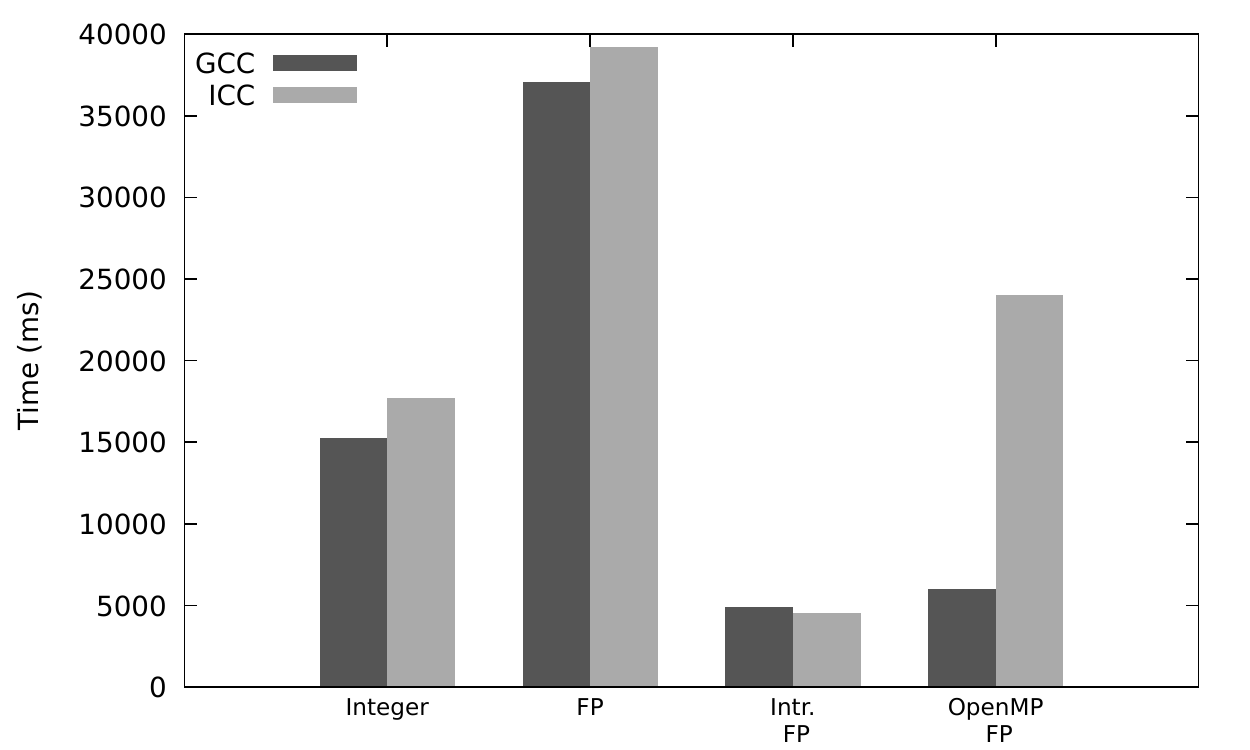}
    }
    \hfill
    \subfigure[On \AAS.]{%
    \includegraphics[width=\longueur\linewidth]{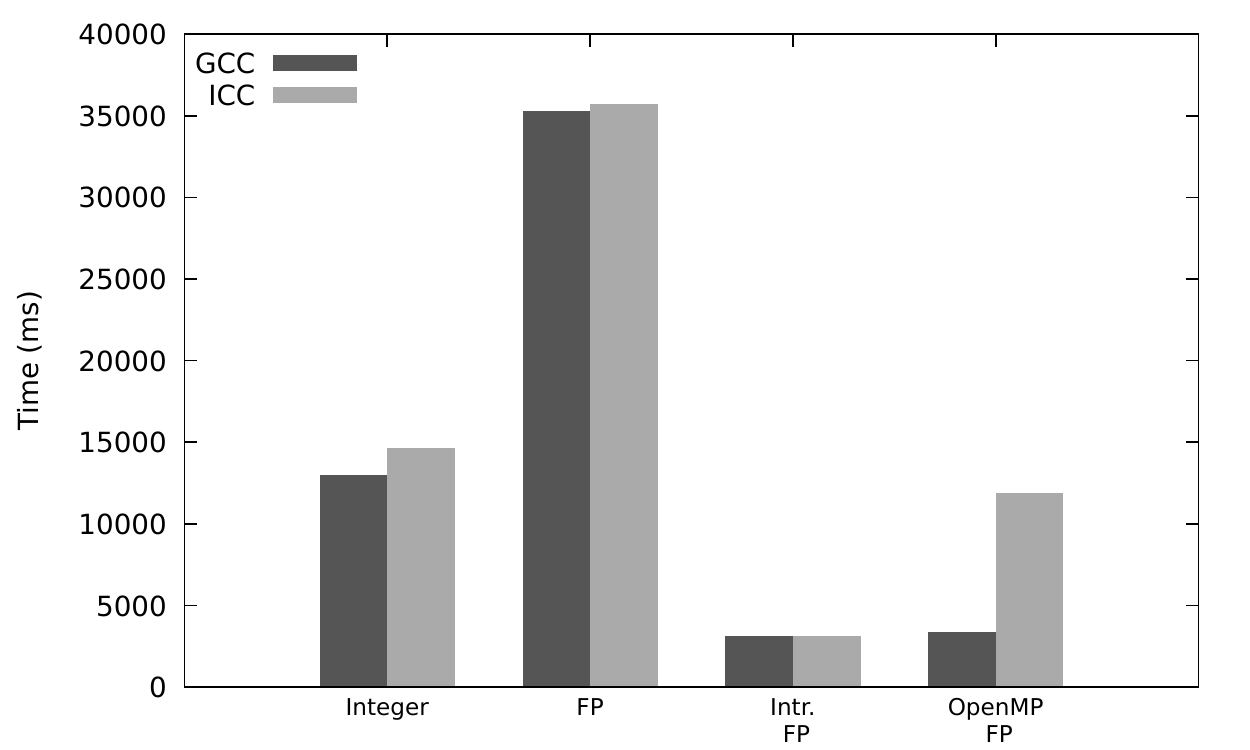}
    }

    \caption{Performance comparison of scalar and SIMD
    polynomial evaluations.  
      \texttt{Integer} denotes the original polynomial evaluation
       by Monagan and coworkers\cite{HM16} with scalar integer-based modular
       arithmetic. 
      \texttt{FP} denotes the use of scalar floating-point modular
      arithmetic for the polynomial evaluation. 
      \texttt{Intr. FP} (respectively \texttt{OpenMP FP})
      denotes the SIMD version of the
      floating-point based modular arithmetic
      using intrinsics (resp. using OpenMP).}\label{f:SIMD4evalnext2}
    
\end{figure}

We can now consider the integration of SIMD modular arithmetic
in our partial polynomial evaluation.
Due to the cost of the conversions 
between integers and floating-points numbers (see
Sect.~\ref{s:ubenchs}), 
we choose to perform the first conversion (from integers to floating-point
numbers) for each value of the $a$ and $m$ vectors once 
 before the evaluation (i.e. just before Line \ref{eval_loop} in
 Algorithm \ref{a:kernel}).
These conversions are performed in-place to save memory.
The reverse conversion (from floating-point
numbers to integers) is only performed once for each reduction result
(i.e. the $c$ value at Line \ref{a:kernel:result}
in  Algorithm \ref{a:kernel}).

Figure \ref{f:SIMD4evalnext2} presents performance results 
for our polynomial evaluation using various modular arithmetic
implementations. 
One can see that the scalar 
floating-point based modular arithmetic
makes our polynomial evaluation about 2.5 times slower
than the
original implementation by Monagan and coworkers\cite{HM16} (using
integer-based modular arithmetic). This is due to the slow FP-based
$\modplus$ implementation (because of its branch instruction: see
Sect.~\ref{s:ubenchs}).

We now consider SIMD intrinsics to integrate 
SIMD FP-based modular arithmetic in our
originally scalar polynomial evaluation (see
Algorithm \ref{a:kernel}).
The resulting SIMD algorithm is written in
Algorithm \ref{a:kernel_simd}.  
First, a reduction has to be computed within the
SIMD vector at the end of the inner loop
(Line \ref{a:kernel_simd:final_reduction}
in Algorithm \ref{a:kernel_simd}), in order to obtain the final scalar 
$c$ value.
Instead of performing a sequential reduction with the scalar
$\modplus$ and its branch instruction,
we use SIMD shuffle
instructions to write a parallel tree-shaped
reduction using only SIMD $\modplus$ operations.
Second, we also have to consider memory
alignement
which can be important for efficient vector loads and stores.
However, the $j$ indices used in the inner loop
(Line \ref{a:kernel_simd:inner_loop}
in Algorithm \ref{a:kernel_simd}) do not lead 
to aligned memory accesses since the successive $J$ values are
not necessarily multiples of the SIMD width.
One could choose to make a copy of the vectors $a$ and $m$ with 
relevant zero padding to ensure aligned memory accesses, but this would
require twice the memory.
In order to obtain good SIMD speedups without
padding, we explicitly decompose this inner loop into three successive 
loops $L1$, $L2$ and $L3$ (not shown in Algorithm \ref{a:kernel_simd}).
$L1$ and $L3$ have an iteration count
lower than the SIMD 
width and are vectorized thanks to explicit masks.
These two loops ensure that the vectorized $L2$ loop 
perform aligned memory accesses
(i.e. its first $j$ index is a multiple of the SIMD width). 
We noticed that the use of such a vectorized reduction and
of such vectorized $L1$ and $L3$ loops offer additionnal 
performance gains when processing a lower number of terms: e.g. up to
29\% for $s = 5 \times 10^4$ terms.
With respect to the original implementation using 
integer-based arithmetic, the resulting polynomial 
evaluation with SIMD intrinsics offers performance gains ranging from
3.13x (resp. 3.89x) with \gcc (resp. \icc) on \AS
to 4.18x (resp. 4.74x) with \gcc (resp. \icc) on \AAS
(see Fig. \ref{f:SIMD4evalnext2}). 

Finally, we also consider using OpenMP vectorization for
the SIMD FP-based modular arithmetic in our
polynomial evaluation.
We rely on the
new \verb+declare reduction+ directive (available
since OpenMP 4.0) to instruct the compiler that
the final reduction (Line \ref{a:kernel_simd:final_reduction}
in Algorithm \ref{a:kernel_simd}) has to be performed
using our modular arithmetic. 
We emphasize that the vectorization is achieved here thanks to
this OpenMP directive (along with the directive which
instructs the compiler to vectorize the loop with a reduction).
Without such directives given by the
programmer, the \gcc and \icc compilers both manage to vectorize
the microbenchmarks presented in Sect.~\ref{s:ubenchs}
(with the
same performance as the OpenMP version),
but both fail to vectorize the polynomial evaluation code 
(because of the required specific reductions). 
As shown in Fig.  \ref{f:SIMD4evalnext2}, the \icc OpenMP
vectorization is inefficient due to the FMA issue (see Sect.
\ref{s:ubenchs}),
whereas the \gcc OpenMP
vectorization offers computation times somewhat slower than the
intrinsic vectorization: 22\% slower on \AS, and 8\% on \AAS. 
It can also be noticed that \gcc currently fails to generate
align memory accesses (despite the use of the \verb+aligned+
OpenMP clause) and SIMD $\modplus$ reductions, as we do with intrinsics.

\smallskip

In conclusion, while the use of scalar FP-based modular
arithmetic lowers the performance of the polynomial evaluation, the
SIMD FP-based modular arithmetic clearly improves its
performance (up to 4.74x).
In the rest of the article, we will rely on the SIMD implementation with
intrinsics, and not on the OpenMP one. This is due to the OpenMP
performance issue with \icc and to the somewhat lower performance of
the SIMD code generated with OpenMP, especially on AVX2 which
still equips  the vast majority of available
CPUs at the time of writing. 
We however emphasize that the performance results of OpenMP with \gcc
on AVX-512 are very promising for the future and show the relevance of
this approach. 

As a last remark, we recall that regarding the microbenchmarks presented in
Sect.~\ref{s:avx-512} 
the AVX-512 $\modtimes$ and $\modplus$
implementations are twice as fast than the AVX2 ones.
Here however for the polynomial evaluation,
the AVX-512 performance is only 1.57x (resp. 1.47x) faster
than the AVX2 one with \gcc (resp. \icc). 
We believe that this is due to the difference in operational intensities.
Indeed, the microbenchmarks
performed in  
Sect.~\ref{s:avx-512} have been intentionally designed to be
compute-bound in order to measure the
number of cycles of the arithmetic operations, and not of the memory accesses.
But the operational intensity of our polynomial evaluation is much
lower: the Hadamard product and the coefficient reduction correspond
to a dot product which is a memory-bound operation in classic
floating-point arithmetic.
More precisely, the floating-point based modular arithmetic requires
9 flop (floating-point operation) for $\modtimes$ (see
Algorithm \ref{a:modtimes}) and 2 flop for
$\modplus$ (see Algorithm \ref{a:modplus}),
 versus 3 memory accesses for each (2 loads and 1 store, without considering
 $u$ and $p$).
This makes our compute kernel (i.e. our polynomial evaluation)
less memory-bound than
a floating-point dot-product,
but the operational intensity of our kernel is not high enough
to make it 
compute-bound: memory accesses are still important
in the kernel performance. These memory accesses
also tend to lower the performance
gain due to the increased
compute power of the AVX-512 SIMD units, with respect to the AVX2
units, 
since there is more stress on memory bandwidth with AVX-512
instructions than with AVX2 ones. 
We will show how to increase this operational intensity and the
polynomial evaluation performance in the
next section.

\section{Increasing the compute efficiency}
\label{s:IncCompEff}

We now focus on the compute efficiency of our SIMD polynomial
evaluation. 
More precisely, we aim to fill at best the pipelined floating-point units and to
minimize the time lost in memory accesses.

\subsection{Multiple dependent evaluations}

We first rely on the consecutive 
powers of $\beta$ used for the successive
evaluations in the matrix method
(see Sect.~\ref{s:matrix_method}).
Hence in Algorithm \ref{a:kernel_simd}, if we consider two consecutive polynomial
evaluations $t$ and $t+1$, the values computed in the $a$ vector
for the evaluation
$t$ are re-used as input for evaluation $t+1$.
But for large $s$ values ($s$ denoting the number of terms, see
Sect.~\ref{s:matrix_method}), the $a$ elements may have been moved
out of the CPU caches.
We hence consider computing multiple evaluations at a
time, and we denote by $T_d$ the number of such (dependent)
evaluations. $T_d$ is an algorithmic constant, known at compile time. 
We can then explicitly avoid storing and reloading data from the vector $a$
to/from memory between these $T_d$ evaluations.
We can also load only once $m$ data from memory for these $T_d$ evaluations. 
Such data reuse increases the operational intensity of our kernel by reducing 
the number of memory accesses. 

However each evaluation depends on the output of the previous one. Even
if some operations
can be performed concurrently (such as the $\modplus$ operation of the
$t$ evaluation and the $\modtimes$ of the $t+1$ evaluation), 
this dependency limits the instruction-level parallelism, and hence
prevents us from filling the pipelines of the floating-point
units. 

\subsection{Multiple independent evaluations}

Therefore, we rewrite the loop over the $T$ evaluations
(Line \ref{eval_loop} in Algorithm \ref{a:kernel}) in order to have 
fully {\it independent} polynomial evaluations.
For this purpose, we adapt the algorithm used by Hu and Monagan\cite{HM16}
to introduce thread-level parallelism on multi-core CPUs 
 (see Sect.~\ref{s:parallel_eval}) in order to introduce here instruction-level
 parallelism in our compute kernel. 
Namely, denoting by $T_i$ the desired number of independent
evaluations (like $T_d$, $T_i$ is an algorithmic constant, known at
compile time), we first precompute 
$\Gamma = [m_1^{T_i},m_2^{T_i},\dots,m_s^{T_i}]$
 using $O(s \log_2 {T_i})$ SIMD multiplications.
We also precompute ${\Lambda}_k = a \circ [m_1^{k+1},m_2^{k+1},\dots,m_s^{k+1}]$ for $0 \le k < {T_i}$
using $O(s T_i)$ SIMD multiplications.
Then, for the computation of the $T$ evaluations we will
first perform the coefficient reductions for
the first $T_i$ evaluations (i.e. on $({\Lambda}_k)_{0 \le k < {T_i}}$),
then the Hadamard product ${\Lambda}_k \gets {\Lambda}_k \circ \Gamma$
(with $0 \le k < {T_i}$)
for the second chunk of $T_i$ evaluations. This will be repeated (coefficient
reductions on the previous $T_i$ evaluations, then Hadamard product for
the next $T_i$ evaluations) until all $T$ evaluations have been processed.

This instruction-level parallelism helps fill the instruction pipelines.
 Moreover the $\modplus$ and
$\modtimes$ operations are now inverted.
Contrary to Algorithm \ref{a:kernel} (Lines \ref{a:kernel:op1}-\ref{a:kernel:op2})
where the second operation depends on the output of the first one,
the second operation now only depends on the input of the first one.
The two operations
can thus be more overlapped, 
hence easing the pipeline filling.

The main drawback of using $T_i$ independent evaluations is
the extra memory requirements.
For each independent evaluation $k$ we have to store an extra copy $\Lambda_k$
of the complete $a$ vector. Moreover, for each
independent evaluation $k$ 
we have to load the $\Lambda_k$ vector from memory and store its update in memory.
The operational intensity is thus only improved for the $\Gamma$ memory
accesses. Therefore, introducing an extra independent evaluation increases
less the operational intensity than introducing an extra dependent evaluation.

There is hence a trade-off between pipeline filling and operational intensity
regarding the numbers of dependent ($T_d$) and independent ($T_i$) evaluations. 
We will thus consider an algorithm where we introduce $T_d$ dependent
evaluations for each of the $T_i$ evaluations, hence computing together
$T_i \times T_d$ evaluations at a time. 
The loop over $T_d$ is chosen as the outer one, and the loop over $T_i$ as the
inner one: this results in better performance than the opposite loop ordering,
which indicates that pipeline filling is here more important than
increasing the operational intensity. 
The optimal values for $T_d$ and $T_i$ depend on the compiler and on the
CPU hardware features,
and these will have to be determined in practice using parameter testing and tuning.
Since the total number of evaluations $T$ 
is not necessarily a multiple of $T_i \times T_d$, the remaining evaluations
are processed first by blocks of $T_i \times N'_d$ evaluations (with
$1 \leq N'_d < T_d$) and then with $N''_d$ dependent evaluations
(with $N''_d < T_i$),
where~ $T \, mod \, (T_i \times T_d) = T_i  \times N'_d + N''_d$\,.

\begin{algorithm}[t]
\begin{algorithmic}[1]
    \State Pre-compute $\Gamma = [m_1^{T_i}, \dots,m_s^{T_i}]$
    and $\Lambda_{k_i} = a \circ [m_1^{{k_i}+1}, \dots,m_s^{{k_i}+1}]$
    for $0 \le k_i < {T_i}$
    \if\HAL1 \Statex \fi (with in-place conversions: 64-bit integers $\rightarrow$ doubles)
    \For{each evaluation $1 \le t \le T$ with step $T_i \times T_d$} 
    \State $i \gets 1;$ $b_{t + k_i + k_d T_i} \leftarrow
    0, \textrm{~for~} (0 \le k_i < T_i ~;~ 0 \le k_d < T_d) $
    \While{$i \leq s$} 
        \State $\bar{c}_{k_i,k_d} \gets 0.0, \textrm{~for~} (0 \le k_i < T_i ~;~ 0 \le k_d < T_d) $
        \State{$J \gets$ \#monomials with same $(d_i, e_i)$}
        \For{$i \le j < i+J $ with step ${\cal V}$} \label{a:kernel_ind_dep:loop_j} 
          \State $ \bar{\Lambda}_{k_i} \gets \Lambda_{k_i}[j~..~j+{\cal V}-1], \textrm{~for~}  0 \le k_i < T_i$ \label{a:kernel_ind_dep:loop_loads} \Comment{SIMD loads\hspace*{3.5cm}~}  
          \State $ \bar{\Gamma} \gets \Gamma[j~..~j+{\cal V}-1] $ \Comment{SIMD load\hspace*{3.5cm}~} 
          \For{$0 \le k_d < T_d$} \label{a:kernel_ind_dep:loop_T_d} 
            \For{$0 \le k_i < T_i$} \label{a:kernel_ind_dep:loop_T_i}
              \State $\bar{c}_{k_i,k_d} \gets \bar{c}_{k_i,k_d} ~\modplus~ \bar{\Lambda}_{k_i}$ \label{a:kernel_ind_dep:op1} \Comment{SIMD coefficient reduction\hspace*{3.5cm}~} 
              \State $\bar{\Lambda}_{k_i} \gets \bar{\Lambda}_{k_i} ~\modtimes~ \bar{\Gamma}$ \label{a:kernel_ind_dep:op2} \Comment{SIMD Hadamard product\hspace*{3.5cm}~} 
            \EndFor 
          \EndFor
          \State $ \Lambda_{k_i}[j~..~j+{\cal V}-1] \gets \bar{\Lambda}_{k_i}, \textrm{~for~}  0 \le k_i < T_i$ \label{a:kernel_ind_dep:loop_stores} \Comment{SIMD stores\hspace*{3.5cm}~}  
        \EndFor
        \For{$0 \le k_d < T_d$} \label{a:kernel_ind_dep:loop_reduction_T_d}
          \For{$0 \le k_i < T_i$} \label{a:kernel_ind_dep:loop_reduction_T_i}
          \State $c_{k_i,k_d} \gets reduce(\bar{c}_{k_i,k_d}, ~\modplus)$ \Comment{$\bar{c}_{k_i,k_d}$ final reduction\hspace*{3.5cm}~}
          \State {\textbf{if} $c_{k_i,k_d} \neq 0.0$ \textbf{then}  convert $c_{k_i,k_d}$ to
          64-bit integer and \if\HAL1 \Statex \hspace*{5,7cm}\fi add 
          $c_{k_i,k_d} x_1^{d_i} x_2^{e_i}$ to bivariate image $b_{t + k_i + k_d T_i}$}
        \EndFor
      \EndFor
      \State{$i \gets i+J$}
    \EndWhile 
  \EndFor 
\end{algorithmic}
\caption{SIMD compute kernel of the matrix method with independent and
      dependent evaluations\label{a:kernel_ind_dep}  (see
      Algorithm \ref{a:kernel} for notations and inputs). 
      $\bar{x}$ denotes the SIMD vector corresponding to variable
      $x$, and $\bar{x}_i$ the {\it i}th SIMD vector. ${\cal V}$ is
      the size of the SIMD vector. 
      }
\end{algorithm}

\begin{figure}[t]
\if\HAL1
\scalebox{0.8}{
\fi 
    \hspace*{1.5cm} \subfigure[Precomputations]{%
    \begin{tabular}{rccccc}
    & & & & & \\
    $\Lambda_0 =$ & $a_1 m_1$ & \dots & $a_i m_i$ & \dots & $a_s m_s$ \\      
    $\Lambda_1 =$ & $a_1 m_1^2$ & \dots & $a_i m_i^2$ & \dots & $a_s m_s^2$ \\   
    $\Lambda_2 =$ & $a_1 m_1^3$ & \dots & $a_i m_i^3$ & \dots & $a_s m_s^3$ \\
    & & & & & \\
    $\Gamma =$ & $m_1^3$ & \dots & $m_i^3$ & \dots & $m_s^3$ \\
    & & & & & \\
    \end{tabular}\label{t:cs:pre}
    }
\if\HAL1
}
\fi 
    \hfill
\if\HAL1
\scalebox{0.8}{
\fi 
    \subfigure[First ${\cal V}$ products of first evaluations]{%
    \begin{tabular}{rcccccccc}
    $\Lambda_0 =$ & \tikzmarkin{a} $a_1 m_1^4$ & \dots $\quad$ & $a_{\cal V} m_{\cal V}^4$ & \dots & $a_s m_s^4$ \\      
    $\Lambda_1 =$ & $a_1 m_1^5$ & \dots $\quad$ & $a_{\cal V} m_{\cal V}^5$ & \dots & $a_s m_s^5$ \\      
    $\Lambda_2 =$ & $a_1 m_1^6$ & \dots $\quad$ & $a_{\cal V} m_{\cal V}^6$ & \dots & $a_s m_s^6$ \\      
    $\Lambda_0 =$ & $a_1 m_1^7$ & \dots $\quad$ & $a_{\cal V} m_{\cal V}^7$ & \dots & $a_s m_s^7$ \\      
    $\Lambda_1 =$ & $a_1 m_1^8$ & \dots $\quad$ & $a_{\cal V} m_{\cal V}^8$ & \dots & $a_s m_s^8$ \\      
    $\Lambda_2 =$ & $a_1 m_1^9$ & \dots $\quad$ & $a_{\cal V} m_{\cal V}^9$ \tikzmarkout{b} & \dots & $a_s m_s^9$ \\      
    & & & & \\
    \end{tabular}\label{t:cs:s1}
    \tikz[overlay,remember picture]{\draw[rounded corners,ultra
    thick,color=red!60]
    ($(a)+(-0.1,\if\HAL0 0.3 \else 0.4 \fi)$) rectangle ($(b)+(0.1,-0.2)$) ;}     
    }
\if\HAL1
}
\fi 

\if\HAL1
\scalebox{0.8}{
\fi 
    \subfigure[Next ${\cal V}$ products of first evaluations]{%
    \begin{tabular}{rcccccccc}
    $\Lambda_0 =$ & $a_1 m_1^4$ & \dots & \tikzmarkin{a} $a_{{\cal V}+1} m_{{\cal V}+1}^4$ & \dots & $a_{2{\cal V}} m_{2{\cal V}}^4$ & \dots & $a_s m_s^4$ \\      
    $\Lambda_1 =$ & $a_1 m_1^5$ & \dots & $a_{{\cal V}+1} m_{{\cal V}+1}^5$ & \dots & $a_{2{\cal V}} m_{2{\cal V}}^5$ & \dots & $a_s m_s^5$ \\      
    $\Lambda_2 =$ & $a_1 m_1^6$ & \dots & $a_{{\cal V}+1} m_{{\cal V}+1}^6$ & \dots & $a_{2{\cal V}} m_{2{\cal V}}^6$ & \dots & $a_s m_s^6$ \\      
    $\Lambda_0 =$ & $a_1 m_1^7$ & \dots & $a_{{\cal V}+1} m_{{\cal V}+1}^7$ & \dots & $a_{2{\cal V}} m_{2{\cal V}}^7$ & \dots & $a_s m_s^7$ \\      
    $\Lambda_1 =$ & $a_1 m_1^8$ & \dots & $a_{{\cal V}+1} m_{{\cal V}+1}^8$ & \dots & $a_{2{\cal V}} m_{2{\cal V}}^8$ & \dots & $a_s m_s^8$ \\      
    $\Lambda_2 =$ & $a_1 m_1^9$ & \dots & $a_{{\cal V}+1} m_{{\cal V}+1}^9$ & \dots & $a_{2{\cal V}} m_{2{\cal V}}^9$ \tikzmarkout{b} & \dots & $a_s m_s^9$ \\      
    & & & & & & \\
    \end{tabular}\label{t:cs:s2}    
    \tikz[overlay,remember picture]{\draw[rounded corners,ultra
    thick,color=red!60]
    ($(a)+(-0.1,\if\HAL0 0.3 \else 0.4 \fi)$) rectangle ($(b)+(0.1,-0.2)$) ;}     
    }
\if\HAL1
}
\fi 
    \hfill
\if\HAL1
\scalebox{0.8}{
\fi 
    \subfigure[First ${\cal V}$ products of next evaluations]{%
    \begin{tabular}{rcccccccc}
    $\Lambda_0 =$ & \tikzmarkin{a} $a_1 m_1^{10}$ & \dots & $a_{\cal V} m_{\cal V}^{10}$ & \dots & $a_s m_s^{10}$ \\      
    $\Lambda_1 =$ & $a_1 m_1^{11}$ & \dots & $a_{\cal V} m_{\cal V}^{11}$ & \dots & $a_s m_s^{11}$ \\      
    $\Lambda_2 =$ & $a_1 m_1^{12}$ & \dots & $a_{\cal V} m_{\cal V}^{12}$ & \dots & $a_s m_s^{12}$ \\      
    $\Lambda_0 =$ & $a_1 m_1^{13}$ & \dots & $a_{\cal V} m_{\cal V}^{13}$ & \dots & $a_s m_s^{13}$ \\      
    $\Lambda_1 =$ & $a_1 m_1^{14}$ & \dots & $a_{\cal V} m_{\cal V}^{14}$ & \dots & $a_s m_s^{14}$ \\      
    $\Lambda_2 =$ & $a_1 m_1^{15}$ & \dots & $a_{\cal V} m_{\cal V}^{15}$ \tikzmarkout{b} & \dots & $a_s m_s^{15}$ \\      
    & & & & \\
    \end{tabular}\label{t:cs:s3}
    \tikz[overlay,remember picture]{\draw[rounded corners,ultra
    thick,color=red!60]
        ($(a)+(-0.1,\if\HAL0 0.3 \else 0.4 \fi)$) rectangle ($(b)+(0.1,-0.2)$) ;}     
    }
\if\HAL1
}
\fi 

\caption{\label{f:cs} Illustration of the execution
 of
Algorithm \ref{a:kernel_ind_dep} 
with $T_i = 3$, $T_d = 2$ and using notations of
Sect.~\ref{s:matrix_method}. 
${\cal V}$ is the size of the SIMD vector. 
For the ease of reading we only represent here the Hadamard products,
but the SIMD coefficient reductions are performed alongside, and the
$\bar{c}_{k_i,k_d}$ final reductions when required. 
First (Fig. \ref{t:cs:pre}), we precompute $\Gamma$ and
$({\Lambda}_k)_{0 \le k < {T_i} = 3}$. Then (Fig. \ref{t:cs:s1}), using $\Gamma$
we compute in SIMD the first
${\cal V}$ products of the first $T_i \times T_d = 3 \times 2$ evaluations
in $({\Lambda}_k)_{0 \le k < 3}$. The next ${\cal V}$ products are performed in
SIMD for the same evaluations (Fig. \ref{t:cs:s2}), and so on for the
remainings of vectors  $({\Lambda}_k)_{0 \le k < 3}$ and $\Gamma$.
Once these first evaluations have been fully computed, we start again with
the next $3 \times 2$ evaluations (Fig. \ref{t:cs:s3}),
until all evaluations have been fully processed.}
\end{figure}

The final version is presented in Algorithm \ref{a:kernel_ind_dep},
along with 
the SIMD programming.
Once $T_i \times T_d$ evaluations have been computed together for some 
monomials, we could choose to iterate over the next $T_i \times
T_d$ evaluations or to iterate over the next 
monomials with same $(d_i, e_i)$ values. 
If we had iterated over the next $T_i \times T_d$ evaluations, we would
have had to store $T_i \times T_d$ SIMD vectors
(all ($\bar{c}_{k_i,k_d})_{0 \le k_i < T_i,~ 0 \le k_d < T_d}$)   
for the coefficient reductions. 
By iterating on the next monomials (as done in
Algorithm \ref{a:kernel_ind_dep}),
$T_i+1$ SIMD loads
(all $(\bar{\Lambda}_{k_i})_{0 \le k_i < T_i}$ and for $\bar{\Gamma}$) and $T_i$
SIMD stores (all $(\bar{\Lambda}_{k_i})_{0 \le k_i < T_i}$) are required. 
As $T_d > 2$ in practice (as confirmed for the optimal configurations in
Sect.~\ref{s:IncCompEff_tests}), 
it is indeed 
preferable to iterate over the next monomials in order
to minimize the number of memory accesses and hence increase the operational
intensity.
This results in the end in an algorithm where we browse all the monomials
with same $(d_i, e_i)$ values 
to compute  $T_i \times T_d$ evaluations at a time.
An illustration of the execution of Algorithm \ref{a:kernel_ind_dep}
is given in Fig. \ref{f:cs}.

\subsection{Loop unrolling}

Loop unrolling\cite{HP17} enables us to remove the exit test at the
end of the loop body and to interleave instructions from successive loop
iterations in order to better fill the pipelines.
The two nested loops over the $T_d$ and $T_i$ evaluations
(Lines \ref{a:kernel_ind_dep:loop_T_d} and
\ref{a:kernel_ind_dep:loop_T_i} in Algorithm  \ref{a:kernel_ind_dep}) 
are hence
completely unrolled thanks to the ``unroll(F)'' pragma of \icc
and to the ``GCC unroll F'' pragma of \gcc (recently introduced in \gcc 8)
to impose an unroll factor of F (F being respectively equal
to $T_d$ and $T_i$).
Similarly, we unroll the third loop over the monomials with same
 $(d_i, e_i)$ values (Line \ref{a:kernel_ind_dep:loop_j} in
 Algorithm  \ref{a:kernel_ind_dep}) by a factor $M$.
We could also have let the
compiler choose which loops to unroll (or not) and determine the
best unroll factors.
This leads to similar performance with \gcc, but to lower performance
with \icc (up to 8.5\% performance loss).
We thus impose our unrollings on the three loops with the
corresponding pragmas. 

Once the three nested loops have been unrolled, we rely on the compiler and
on the
out-of-order execution of the processor to schedule at best the instructions
to fill the pipelines and to overlap the memory accesses. 
Other loops over $T_d$ and/or $T_i$ evaluations (Lines
\ref{a:kernel_ind_dep:loop_loads}, \ref{a:kernel_ind_dep:loop_stores},
\ref{a:kernel_ind_dep:loop_reduction_T_d},
\ref{a:kernel_ind_dep:loop_reduction_T_i}, 
 in Algorithm \ref{a:kernel_ind_dep}) are also similarly unrolled.

\subsection{Performance results}
\label{s:IncCompEff_tests}

\begin{figure}[t]

    \centering
    \subfigure[With \gcc on \AS.]{%
    \includegraphics[width=\longueur\linewidth]{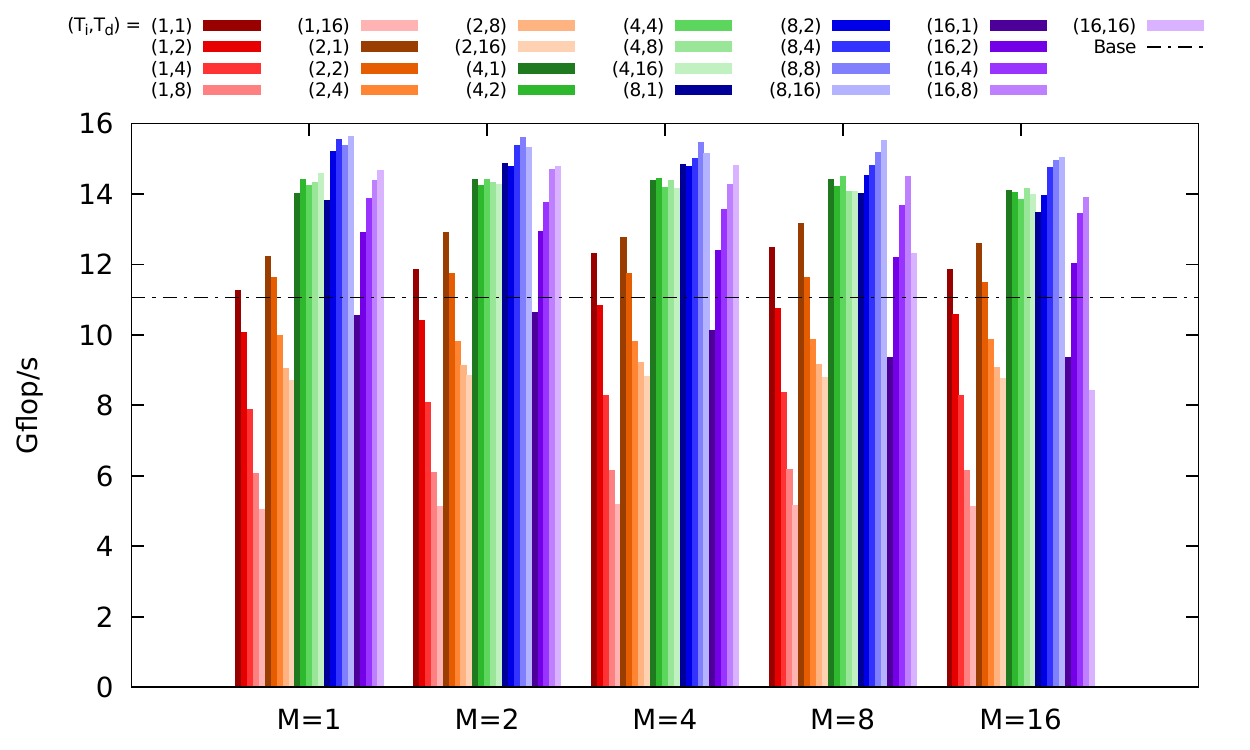}
    }
    \hfill
    \subfigure[With \icc on \AS.]{%
    \includegraphics[width=\longueur\linewidth]{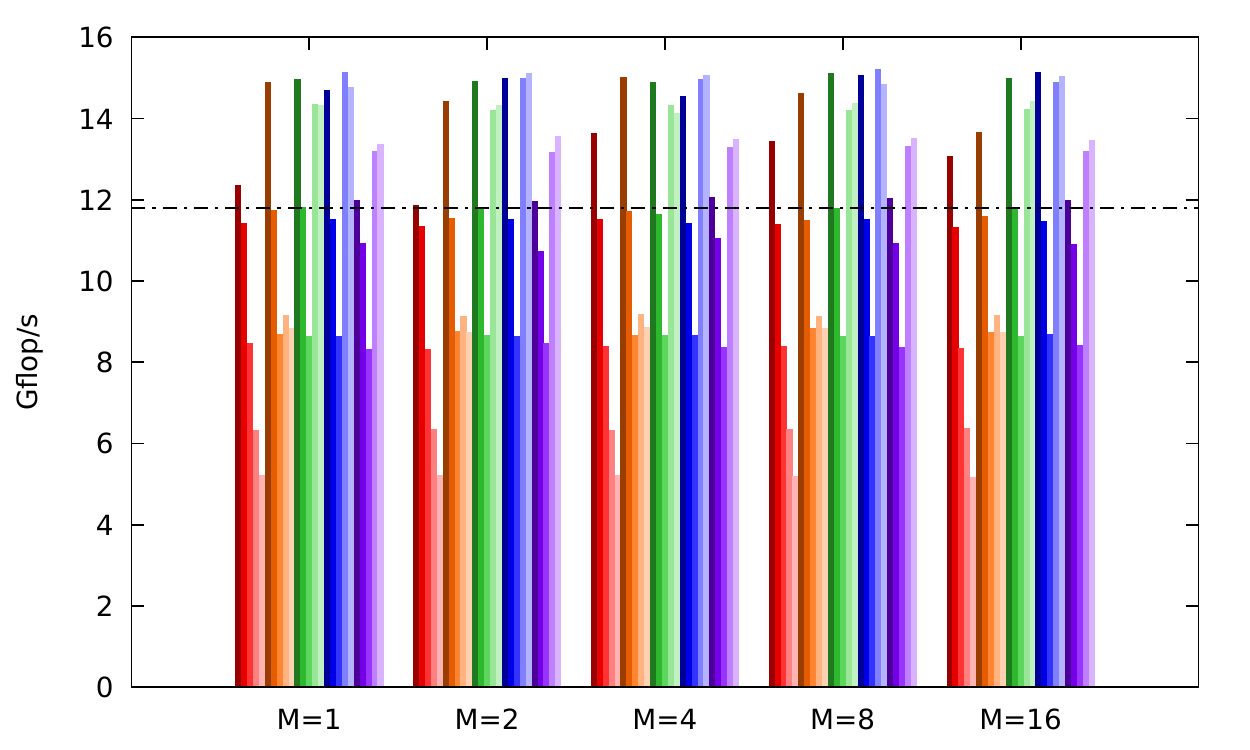}
    }

    \subfigure[With \gcc on \AAS.]{%
    \includegraphics[width=\longueur\linewidth]{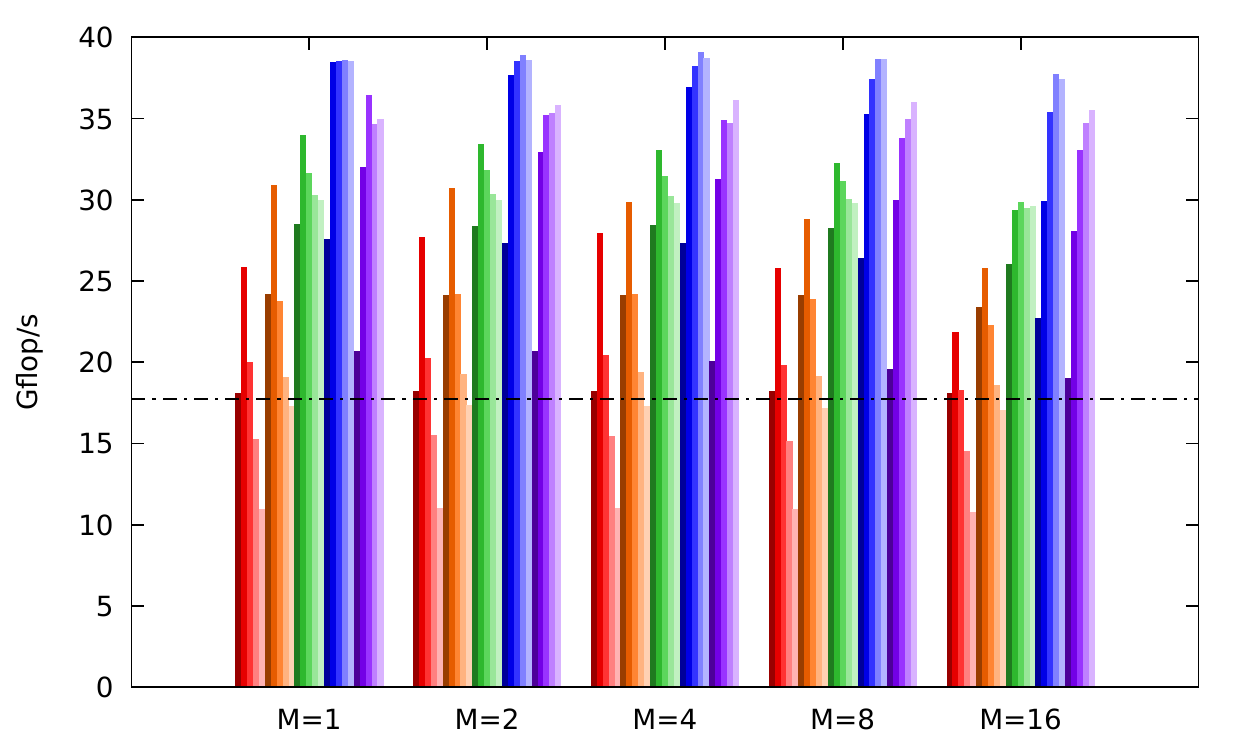}
    }
    \hfill  
    \subfigure[With \icc on \AAS.]{%
    \includegraphics[width=\longueur\linewidth]{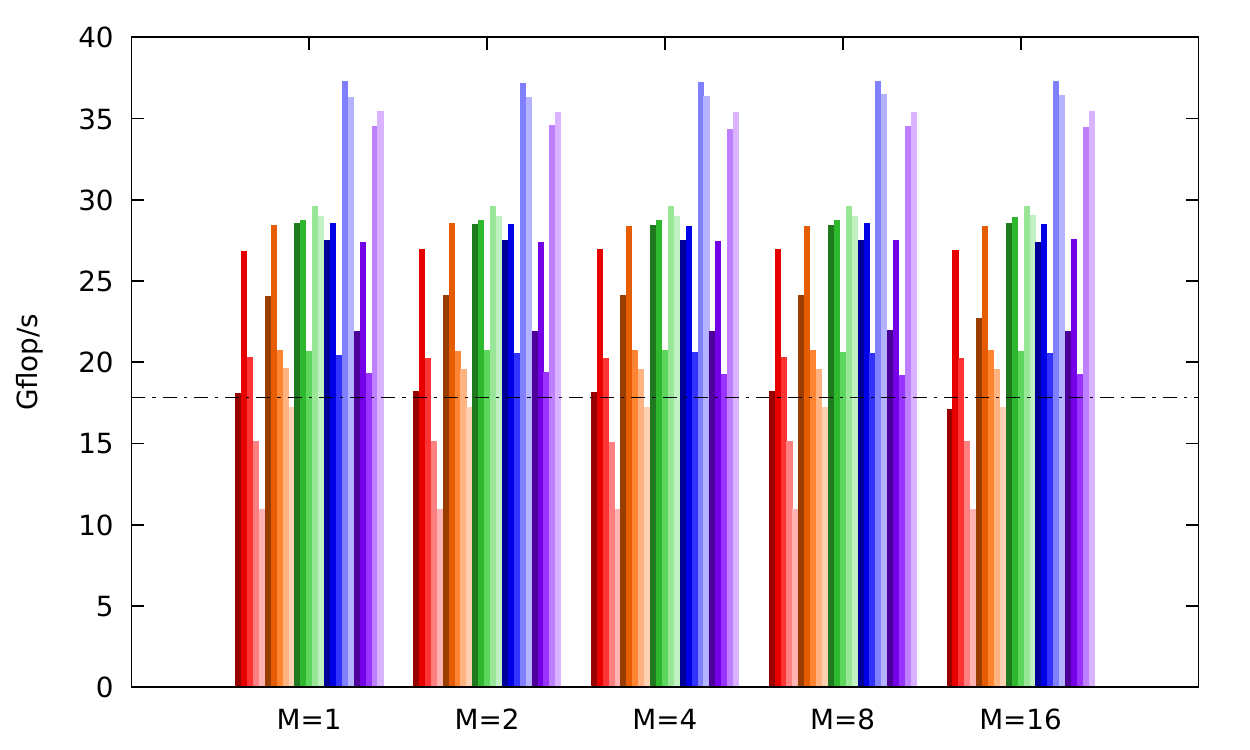}
    }

    \caption{Performance results for all possible ($T_i$, $T_d$, $M$)
    configurations. 
     }\label{f:tuning}
\end{figure}

Using the test platforms (server and compiler)
described in Table \ref{t:platforms}, we present in Figure \ref{f:tuning}
the performance results for all possible configurations
for ($T_i$, $T_d$, $M$), each value ranging in 1,2,4,8,16.
We also indicate the performance of the \texttt{Base} SIMD code corresponding to
the version obtained with the SIMD intrinsics only (as presented in
Sect.~\ref{s:SIMD}). 
The performance varies significantly depending on the ($T_i$, $T_d$, $M$)
values, especially on $T_i$ and $T_d$, which shows the relevance of these
parameters. The performance impact of $M$ is lower
but can still reach 11\% for some ($T_i$, $T_d$) configurations.
The best configurations are the following.
\begin{itemize}
\item ($T_i=8$, $T_d=16$, $M=1$) with \gcc on \AS: 15.65 Gflop/s, and 42\% of performance gain
over the \texttt{Base} SIMD code. 
\item ($T_i=8$, $T_d=8$, $M=8$) with \icc on \AS:  15.21 Gflop/s, and 29\% of performance gain
over the \texttt{Base} SIMD code. 
\item ($T_i=8$, $T_d=8$, $M=4$) with \gcc on \AAS: 39.11 Gflop/s, and 121\% of performance gain
over the \texttt{Base} SIMD code.  
\item ($T_i=8$, $T_d=8$, $M=1$) with \icc on \AAS: 37.31 Gflop/s, and 109\% of performance gain
over the \texttt{Base} SIMD code.  
\end{itemize}

As determining the theoretical peak performance of modern CPUs becomes
more and more complicated\cite{Dol18}, we use the BLAS DGEMM routine of the
Intel MKL\footnote{See: \url{https://software.intel.com/en-us/mkl}}
to estimate the single-core double-precision peak performance
at 45 Gflop/s on \AS and 101 Gflop/s on \AAS.
Moreover, we
 can only reach
61\% of the peak performance since there are only 2 FMAs out of the 9
floating-point instructions required
for $\modtimes$ and $\modplus$. We manage hence to reach
57\% and 63\% of the attainable single-core peak performance, 
respectively on \AS and on \AAS.

\begin{table}[t]
\caption{Performance comparison between our best version and the
reference implementation using integer-based arithmetic.\label{t:final_cmp}}
\if\HAL1
{\small
\begin{center}
\fi
  \begin{tabular}{c|c|c|c|c}
\if\HAL0
     \multirow{2}{0.1\linewidth}{\centering Server} &
     \multirow{2}{0.1\linewidth}{\centering Compiler} &
     \multirow{2}{0.25\linewidth}{\centering Reference scalar
     integer-based version (time in ms)} &
     \multirow{2}{0.3\linewidth}{\centering SIMD FP-based version
     with improved compute efficiency (time in ms)} &
     \multirow{2}{0.1\linewidth}{\centering Gain} \\
\else
     \multirow{3}{0.1\linewidth}{\centering Server} &
     \multirow{3}{0.1\linewidth}{\centering Compiler} &
     \multirow{3}{0.22\linewidth}{\centering Reference scalar
     integer-based version (time in ms)} &
     \multirow{3}{0.3\linewidth}{\centering SIMD FP-based version
     with improved compute efficiency (time in ms)} &
     \multirow{3}{0.07\linewidth}{\centering Gain} \\
    & & & & \\
\fi 
    & & & & \\
     \hline
     \AS   & \gcc & 15262 & 3482 & 4.4x\\
     \AS   & \icc & 17704 & 3671 & 4.8x\\
     \AAS  & \gcc & 12984 & 1411 & 9.2x \\
     \AAS  & \icc & 14638 & 1476 & 9.9x \\
\end{tabular}
\if\HAL1
\end{center}
}
\fi
\end{table}

In the end, as shown in Table \ref{t:final_cmp}, we manage to reach
speedups of almost 5x and 10x (respectively on \AS and on \AAS) on one
CPU core over
the reference original polynomial evaluation with scalar integer-based
modular arithmetic. 
It can be noticed that Monagan and coworkers already used to process
two (dependent) evaluations at a time to increase the operational 
intensity for some variants of the polynomial evaluation.
However for the variant studied in this article 
(see Algorithm \ref{a:kernel}), 
which is the fastest one, no performance gain is obtained by
processing two evaluations at a time with the original scalar code.
Such divergence with respect to the gains obtained in Fig. \ref{f:tuning}
can be explained by
the lack of other optimizations (multiple independent evaluations,
loop unrollings) as well as by the differences
in the modular arithmetic implementation between the original
integer-based version 
by Monagan and coworkers 
and the floating-point based 
version of this article. 

\subsection{Without extra memory requirements}
\label{s:wo_NIB}

\begin{figure}[t]

    \centering
    \includegraphics[width=0.7\linewidth]{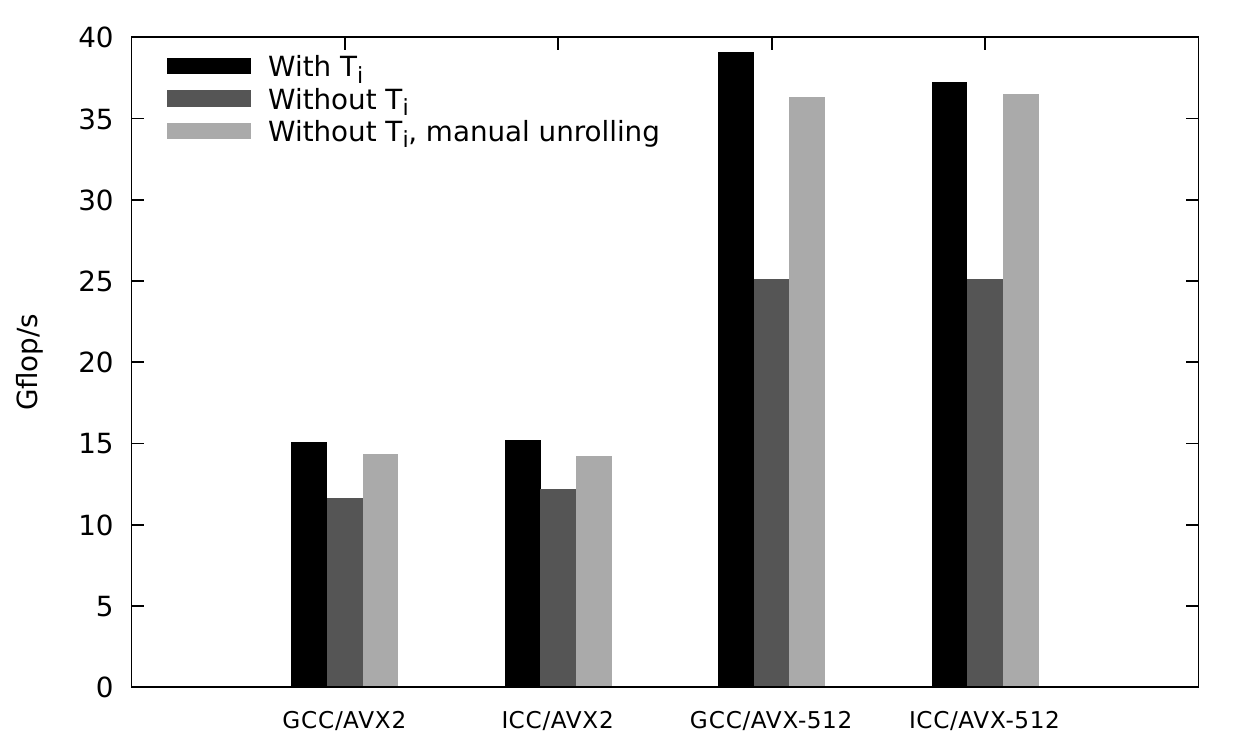}
\caption{Performance comparison between best configuration with
extra memory for multiple independent evaluations (\texttt{with $T_i$}) and best
configurations without extra memory for multiple independent
evaluations (\texttt{without $T_i$}), with and without manual loop unrolling.
\label{f:cmp_wo-NIB}} 
\end{figure}

One drawback of using multiple independent evaluations is
the significant memory overhead: the complete $a$ vector has to
be duplicated for each extra independent evaluation.
We therefore investigate here the best attainable performance without any
extra independent evalution.

We first implement a code
without multiple independent evaluation, and tune the $T_d$ and $M$ parameters
for this code via extensive benchmarks (as in Sect.~\ref{s:IncCompEff_tests}). 
Figure \ref{f:cmp_wo-NIB} shows for each test platform the performance drop
obtained for this code (referred to as \texttt{Without $T_i$}) with respect to
the best version obtained in Sect.~\ref{s:IncCompEff_tests} (referred to as
\texttt{With $T_i$}).  
The performance drop is important here (up to 36\%),
due to the lower number of
independent instructions to fill the pipelines. 

We then introduced manual loop unrolling, using
preprocessor macros to ease and automate the tedious code writing.
We also manually group all arithmetic instructions.
This way, we provide all arithmetic instructions
for the computation of $T_d$ dependent evaluations and $M \times {\cal V}$
monomials to the compiler and to the
out-of-order execution of the processor, so that  
these can be scheduled at best to fill the pipelines.
One can see that this new version (referred to as: \texttt{Without
$T_i$, manual unrolling}; and after tuning of its $T_d$ and $M$ parameters)
greatly 
reduces the performance drop
with respect to the best version obtained in Sect.~\ref{s:IncCompEff_tests}
(\texttt{With $T_i$}). 
This way, we can reach 
95\% (\gcc) 
and 94\% (\icc) 
on \AS 
and 93\% (\gcc) 
and 98\% (\icc) 
on \AAS of the best
attainable performance (\texttt{With $T_i$}).
At the price of non-negligeable development efforts, we can thus
obtain, without introducing extra memory, almost the same
performance of our best versions with multiple independent evaluations. 

It can also be noticed that the performance impact of $M$ is here much
more important than in Sect.~\ref{s:IncCompEff_tests} (detailed tests not
shown).
Such manual loop unrolling and instruction grouping
have also been tried on the best version obtained
in Sect.~\ref{s:IncCompEff_tests} (with multiple independent evaluations): 
this however only offers up to 5.1\% performance gain
for such code.
In our opinion, this does not justify the
manual unrolling development effort 
when using multiple independent evaluations.

\section{Conclusion}
\label{s:conclusion}

In this article, we have first justified the choice of a modular
multiplication algorithm relevant for HPC and SIMD computing.
We have ensured the correct use of an optimized AVX2 implementation
(regarding a potential issue with signed zeros and
the \texttt{blendv\_pd} intrinsic)  
and we have presented its AVX-512 version. 
This floating-point (FP) based algorithm with FMAs (fused
multiply-adds) enables us to obtain SIMD speedups of up to 3.7x on
AVX2, and up to 7.2x on AVX-512, which validates its efficiency. 
With respect to a reference (scalar) 
integer-based modular arithmetic, the performance gains are similar
for our SIMD FP-based modular multiplication and for the corresponding SIMD
FP-based modular addition. 
As all current desktop and HPC processors have SIMD units,
we believe that
such SIMD FP-based modular arithmetic should from now on
be used instead of the scalar ones. 
Using OpenMP for their SIMD programming
turned out to be a very promising approach on the new AVX-512 units
(with \gcc), due to its 
very relevant performance-programmability trade-off.
 Currently, we still rely on
intrinsics programming for best performance and performance
portability among compilers.

In a second part, we have focused on the partial polynomial evaluation
which is a key computation in Computer Algebra. 
We have rewritten this algorithm in order to introduce multiple
independent and dependent evaluations. These enable us, along with
loop unrolling, to fill at best the pipelined floating-point
units of the CPU and to minimize the time lost in memory accesses.
Combined with SIMD computing, we achieve speedups up to almost 5x
on AVX2 and up to almost 10x on AVX-512 with respect to the reference
implementation of the 
polynomial evaluation. 
Moreover, using manual loop unrolling we manage to closely reach
such performance
gains without extra memory requirements.

In the future, we plan to integrate our efficient polynomial
evaluation on one CPU core in the multi-core parallel implementation
of Monagan and coworkers\cite{HM16,MT18}, and to study the performance
impact on polynomial factorizations and polynomial greatest common divisor
computations.
We also believe that GPUs may be well suited to further accelerate
our polynomial evaluation thanks to their higher compute power and
memory bandwidth. We emphasize that our FP-based modular arithmetic will be very
relevant for the GPU FMA SIMD units, and will offer a direct and
efficient implementation of modular arithmetic on GPUs. 
We may also investigate using a few less bits for our prime $p$
in order to decrease 
the number of reductions as done for example  with error-free
transformations in linear algebra\cite{JG10}.

\section*{Acknowledgments}

The authors would like to thank the master in computer science at
Sorbonne Universit\'e, especially N. Picot and P. Cadinot, 
 for administering 
 and providing access to the compute servers.
They also thank Professor S. Graillat (Sorbonne Universit\'e)
for helpful discussions on
error-free transformations.

\if\HAL1
\bibliographystyle{unsrt}
\fi

\bibliography{article}%

\end{document}